%% file: main_text.tex
\documentclass[
superscriptaddress,
amsmath,
amssymb,
achemso,
aps,
prl,
twocolumn,
]{revtex4-2}

\usepackage{braket}
\usepackage{graphicx}
\usepackage[colorlinks=true, allcolors=blue]{hyperref}
\usepackage{gensymb}        
\usepackage{derivative}     
\usepackage{upgreek}        
\usepackage{float}  
\usepackage{xcolor}
\usepackage{footmisc}
\usepackage{siunitx}
\usepackage{calrsfs}
\usepackage{mathrsfs}
\usepackage{booktabs}
\usepackage{mathtools}
\usepackage{physics}
\usepackage{ulem}           
\usepackage{chngcntr}       
\usepackage{cleveref}       

\crefname{appendix}{Supplementary Note}{Supplementary Notes}
\Crefname{appendix}{Supplementary Note}{Supplementary Notes}

\AtBeginDocument{\RenewCommandCopy\qty\SI} 

\makeatletter
\def\maketitle{
    \@author@finish
    \title@column\titleblock@produce
    \suppressfloats[t]
}
\makeatother

\newcommand{\ca}{$\left.^{40}\mathrm{Ca}^{+}\right.$}

\newsavebox\mcFcontent
\savebox\mcFcontent{$\mathcal{F}$}

\begin{document}

\title{Persistent Quantum-Enhanced Frequency Sensing with T\textsuperscript{-3/2} Scaling}

\author{Clayton Z. C. Ho}
\thanks{These authors contributed equally to this work.}
\author{Hao Wu}
\thanks{These authors contributed equally to this work.}
\author{Grant D. Mitts}
\thanks{These authors contributed equally to this work.}
\author{Joshua A. Rabinowitz}
\author{Eric R. Hudson}
\affiliation{Department of Physics and Astronomy, University of California Los Angeles, Los Angeles, CA, USA}
\affiliation{Challenge Institute for Quantum Computation, University of California Los Angeles, Los Angeles, CA, USA}
\affiliation{Center for Quantum Science and Engineering, University of California Los Angeles, Los Angeles, CA, USA}

\date{\today}

\begin{abstract}
Quantum sensing uses nonclassical states to improve measurement sensitivity, but the same states that provide metrological gain also decohere more rapidly.
This limits the usable interrogation time and, in practice, often precludes improvement in \textit{ultimate} sensitivity -- realized useful quantum advantage has consequently remained rare.
Here, we restore persistent quantum advantage by embedding Fock-state enhancement within a quantum heterodyne (Qdyne) protocol, decoupling sensitivity from the decoherence-limited interrogation time $\tau$.
Measurements are acquired at short $\tau$ where the quantum-enhanced gain is optimal, while precision accumulates with the total measurement time $T$.
Demonstrated on the motional mode of a trapped \ca\ ion, we observe quantum-enhanced precision that persists to measurement times seven orders of magnitude beyond the dephasing limit, scaling as $T^{-3/2}$ with no indication of saturation.
Using the $n=3$ Fock state, we reach a frequency precision of \qty{0.5}{\micro\Hz} relative to an \qty{86}{\MHz} carrier, achieving a fractional precision $\sim 6\times10^{-15}$.
This represents a quantum-enhanced gain of \qty{7.1(1.0)}{\dB} over the $n=0$ state, in agreement with Fisher-information predictions.
This is the first demonstration of Qdyne beyond solid-state spin-defects.
Further, by using the quantum harmonic oscillator to perform frequency mixing, we extend operation beyond~\qty{1}{\GHz}, two orders of magnitude above the ceiling of pulsed dynamical-decoupling implementations.
These results recover quantum advantage at the long timescales required to improve \textit{ultimate} sensitivity, with direct implications for nanoscale NMR and quantum logic spectroscopy.
\end{abstract}

\maketitle

Measurements restricted to classical resources are bounded by the standard quantum limit (SQL), the precision floor imposed by quantum projection noise.
A core aim of quantum sensing is to circumvent this limit through the use of nonclassical states~\cite{CavesSeminal1981}, enabling ever greater sensitivities for applications such as quantum illumination~\cite{QuantumIllumination2010}, frequency metrology~\cite{SqueezedClockLattice2025JunYe,SqueezeAmplificationClockVuletic2025}, and optical magnetometry~\cite{SqueezedLightMagnetometry}.
On quantum harmonic oscillator (QHO) platforms, this enhancement has been pursued by exploiting the distinct properties of different nonclassical states.
Notably, Fock states have been shown to be optimal for phase-insensitive displacement measurements~\cite{Wolf2019, Deng2024}, while Fock state superpositions have been used to demonstrate optimal frequency sensitivity~\cite{McCormick2019,KellerQHOSpecanal2021}.
Similarly, Schr\"{o}dinger-cat states have been used for noise spectroscopy~\cite{Milne2019} and single-quanta quantum logic spectroscopy~\cite{Hempel2013,NISTqlsBichromatic2022}, and GKP states have been used for multiparameter sensing~\cite{GKPSensingValahu2024}.

Despite these achievements, quantum-enhanced measurements have rarely improved \textit{ultimate} sensitivity in practice -- squeezed-light gravitational-wave interferometry being a notable exception~\cite{LIGOSqueezedGWDetection2019,LIGOBroadbandSqueezing2023}.
This is because nonclassical states, while offering sub-SQL scaling, suffer accelerated decoherence that can outpace their metrological gain and eliminate any improvement in ultimate sensitivity -- this is well established for multi-qubit GHZ states \cite{GHZClockDecoherence1997, ShajiCaves2007, GHZDecoherenceBlatt2011, QuantumMetrologyNonclassical2018} and certain bosonic states \cite{Myatt2000, Turchette2000}.
The archetypal example is frequency sensing with Fock state superpositions $\ket{\psi} = (\ket{0} + \ket{N})/\sqrt{2}$: enhancement grows as $N$~\cite{McCormick2019} against a dephasing rate that grows as $N^{2}$ \cite{Myatt2000, Turchette2000}, and ultimate sensitivity does not improve; other states and noise models lead to more favorable situations~\cite{Wolf2019}.
Restoring persistent quantum advantage at long timescales therefore requires a protocol that reduces the dependence of precision on interrogation time.

In this work, we demonstrate that a quantum heterodyne protocol (Qdyne~\cite{RetzkerQdyne2017}, also known as CASR~\cite{DegenCASR2017}) achieves exactly this decoupling, and that the combination of Qdyne with quantum enhancement yields persistent metrological gain that improves \textit{ultimate} sensitivity, limited only by reference clock stability~\cite{DegenCASR2017,RetzkerQdyne2017}.
Qdyne, a method developed to acquire nanoscale nuclear magnetic resonance (NMR) spectra with nitrogen-vacancy (NV) centers in diamond, comprises a series of phase-sensitive measurements coherently synchronized by a reference clock and processed by Fourier analysis.
This decouples the decoherence of the quantum resource, which bounds the interrogation time $\tau$, from the total measurement time $T$ over which frequency information accumulates and which can be extended arbitrarily.

We first demonstrate a wideband implementation of Qdyne on the bosonic motional mode of a trapped \ca~ion and reconstruct multi-signal spectra.
To our knowledge, this is the first implementation of Qdyne beyond solid-state spin-defects~\cite{CASRonHBN2023Munich, CASRonSiC2023Tianjin}, establishing its relevance beyond spin sensors and enabling high-resolution sensing on trapped ions.
Moreover, our implementation eliminates a principal bandwidth restriction of Qdyne: the pulsed dynamical-decoupling sequences used in previous demonstrations, which tie the frequency range to the achievable $\pi$-pulse rate, and thus limit operation to tens of \unit{\MHz}~\cite{RetzkerQdyne2017,DegenCASR2017,RetzkerResolution2019,WalsworthCASR2018,NanoNMRReviewBucher2022,NarrowBandwidthCDD2017Retzker}.
Instead, we use motional Raman transitions~\cite{Wu2025,Wu2025super,SubharmonicsEGGS2025} to heterodyne the target signal to the QHO resonance, such that frequency conversion is directly effected by the Hamiltonian, enabling Qdyne operation beyond \qty{1}{\GHz} -- two orders of magnitude beyond the ceiling of pulsed dynamical-decoupling implementations~\cite{RetzkerQdyne2017,DegenCASR2017,WalsworthCASR2018}.
We then implement quantum enhancement with nonclassical Fock states~\cite{Wolf2019}, demonstrating a \qty{7.1(1.0)}{\dB} gain in frequency precision with the $n=3$ state over the $n=0$ state.
Applied to an unknown signal, quantum-enhanced Qdyne achieves a frequency precision $\sigma_{\delta}/(2\pi) = \qty{0.5}{\micro \Hz}$, corresponding to a fractional precision $\sim 6\times 10^{-15}$ when referenced to the \qty{86}{\MHz} carrier.
In what follows, we describe the experimental apparatus and the generation of motional Raman transitions, introduce the motional Raman Qdyne protocol, demonstrate wideband reconstruction of single- and multi-tone spectra at \qty{1}{\GHz}, and show that the integration of Fock-state enhancement delivers metrological gain that persists to arbitrary measurement times -- improving the ultimate sensitivity.

Demonstrations are conducted on the radial motional mode of a trapped \ca~ion with secular frequency $\omega_{0}/(2\pi) \approx \qty{1.285}{\MHz}$.
The ion is trapped in a cryogenic $\approx \qty{4.5}{\K}$ linear Paul trap with radial ion-electrode distance $r_{0} \approx \qty{550}{\um}$.
Spin--motion coupling for motional state manipulation is provided by sideband (Jaynes--Cummings-type) interactions on the $\{\ket{^2\mathrm{S}_{1/2}, m_J = -1/2}, \ket{^2\mathrm{D}_{5/2}, m_J = -5/2}\} \equiv \{\ket{\downarrow}, \ket{\uparrow}\}$ optical qubit manifold of \ca{}.
Doppler cooling followed by resolved-sideband cooling on the optical qubit transition prepares the ion near the ground motional state with $\bar{n} = 1.4(7) \times 10^{-2}$ on the target mode; all other motional modes are cooled to $\bar{n} < 5\times10^{-2}$ to suppress decoherence from spectator modes and anharmonic mode-mode couplings~\cite{SpectatorModeRabi025,MixedSpeciesAnharmonicityJPHome2011}.
State detection is achieved via state-dependent fluorescence~\cite{Christensen2020}: population in $\ket{\uparrow} = \ket{^2\mathrm{D}_{5/2}}$ is dark, while population in $\ket{\downarrow} = \ket{^2\mathrm{S}_{1/2}}$ scatters photons on the $\left.^2\mathrm{S}_{1/2}\right. \rightarrow \left.^2\mathrm{P}_{1/2}\right.$ transition during a \qty{750}{\us} readout window.

All motional Hamiltonians are generated by coupling voltages from an arbitrary waveform generator (AWG) to trap electrodes (\Cref{SI:methods_signal_gen}).
Voltages $V_d$ ($V_q$) applied at frequencies $\omega_d$ ($\omega_q$) with relative phases $\phi_\textrm{AWG} = \pi$ ($0$) between the two AWG channels produce dipolar (quadrupolar) fields with lab-frame Hamiltonians 
\begin{align}   \label{eq:methods_dipole_x_ham}
    \hat{H}_{\text{d}} &= -\frac{\lvert e \rvert V_d}{r_0} \, x_0 \left( \hat{a} + \hat{a}^{\dagger} \right) \, \cos{\!\left(\omega_d t + \phi_d \right)}, \\
    \hat{H}_{\text{q}} &= -\frac{\lvert e \rvert V_q}{2 r_0^{2}} x_0^2 \left( \hat{a} + \hat{a}^{\dagger} \right)^2  \, \cos{\!\left(\omega_q t + \phi_q \right)},
\end{align}
where $x_0 = \sqrt{\hbar/(2 m\omega_0)}$ is the length-scale of the QHO, $\omega_0$ the QHO secular frequency, $\lvert e \rvert$ the magnitude of the electron charge, and $\phi_d ,\, \phi_q$ the phases of the dipole and quadrupole signals, respectively.
For convenience, we define $\Omega_d = \left|e\right| V_d x_0 / (\hbar r_0)$ ($\Omega_q = \left|e\right| V_q x_0^2/ (2 \hbar r_0^2)$) as effective coupling strengths for the dipole (quadrupole) tones.

Simultaneous application of a dipole tone at frequency $\omega_d + \delta$ with phase $\phi_d$ and two quadrupole tones at $\omega_d \pm \omega_0$ with phases $\phi_b ,\, \phi_r$ excites motional Raman transitions that interfere to produce a phase-sensitive displacement~\cite{Wu2025}
\begin{equation}    \label{eq:qvsa_displacement}
    \alpha = \, \Omega_q \Omega_d \frac{ i \omega_0 e^{i \phi_-}}{\omega_d^2 - \omega_0^2} \, \frac{\sin(\delta t - \phi_d + \phi_+) + \sin(\phi_d - \phi_+)}{\delta} \,,
\end{equation}
where $\phi_{\pm} = (\phi_b \pm \phi_r)/2$.
The displacement therefore oscillates with the relative phase $\phi_d - \phi_+$ between the dipole and quadrupole tones.
Throughout this work, signals under test are applied as dipole tones, while quadrupole tones serve as probes.
Stark shifts are absorbed into the secular frequency $\omega_0$ via in-situ calibration, and residual off-resonant excitation is suppressed by shaping pulses with a tapered-cosine (Tukey) envelope.

\begin{figure*}[t!]
    \centering
    \includegraphics[width=\textwidth]{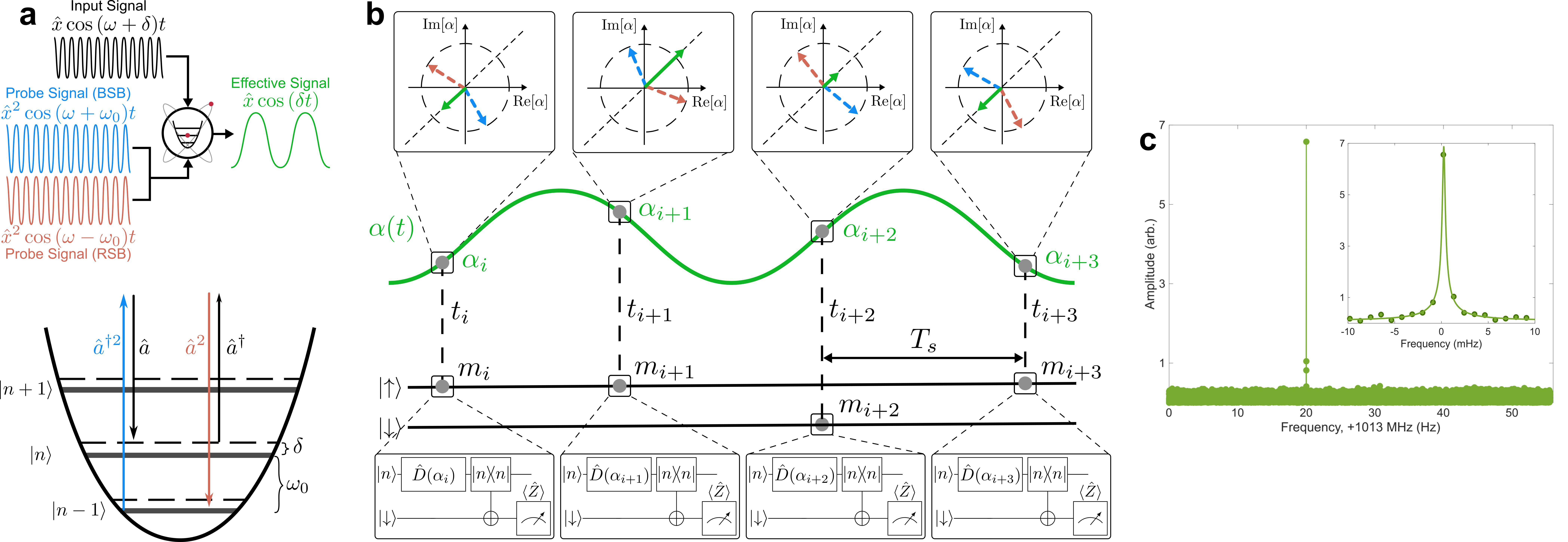}
    \caption{
    \textbf{Quantum-enhanced Qdyne on a trapped-ion quantum harmonic oscillator (QHO).}
    \textbf{(a)} Quantum frequency downconversion via motional Raman excitation.
    Quadrupole probe tones at $\omega_d \pm \omega_0$ (blue, red) mix with a dipole signal tone at $\omega_d + \delta$ (black), driving interfering two-phonon Raman pathways that produce a displacement $\hat{D}{(\alpha)}$ whose amplitude oscillates at a baseband detuning $\delta$ (green).
    \textbf{(b)} Stroboscopic sampling of the downconverted signal for Qdyne.
    The QHO probes the signal with measurements of duration $\tau$, repeated with sample period $T_s$; phase-tracking of tones against a fiducial reference ensures mutual coherence of successive measurements.
    At each instance $t_i$, the red- and blue-sideband Raman contributions (top, dashed arrows) interfere to produce a net displacement $\alpha_i$ (top, green arrow) set by the instantaneous signal phase (\cref{eq:qvsa_displacement_casr}).
    The motional state is converted to a binary outcome $m_i$ by projection onto the prepared Fock state $\ket{n}$ and mapped onto spin for detection via state-selective fluorescence (bottom, circuits) (\Cref{SI:sec_methods}).
    \textbf{(c)} Wideband Qdyne demonstration.
    A dipole tone at $\omega_d/(2\pi) = \qty{1013}{\MHz}$ is heterodyned to $\delta/(2\pi) = \qty{10}{\Hz}$ and sampled with $T_s = \qty{9}{\ms}$ over $T = \qty{900}{\s}$ ($N = 10^{5}$ samples).
    A discrete Fourier transform (DFT) of the measurement record $\{m_i\}_{1}^{N}$ recovers the signal as a peak within the Nyquist-limited bandwidth $1/(2T_s) = \qty{55.6}{\Hz}$; the peak appears at $2\delta/(2\pi) = \qty{20}{\Hz}$ as overlap readout (\cref{eq:fock_overlap_prob}) depends on $\lvert\alpha\rvert^{2}$.
    \label{fig:1_technique_overview}
    }
\end{figure*}

\begin{figure*}[t!]
    \centering
    \includegraphics[width=0.8\textwidth]{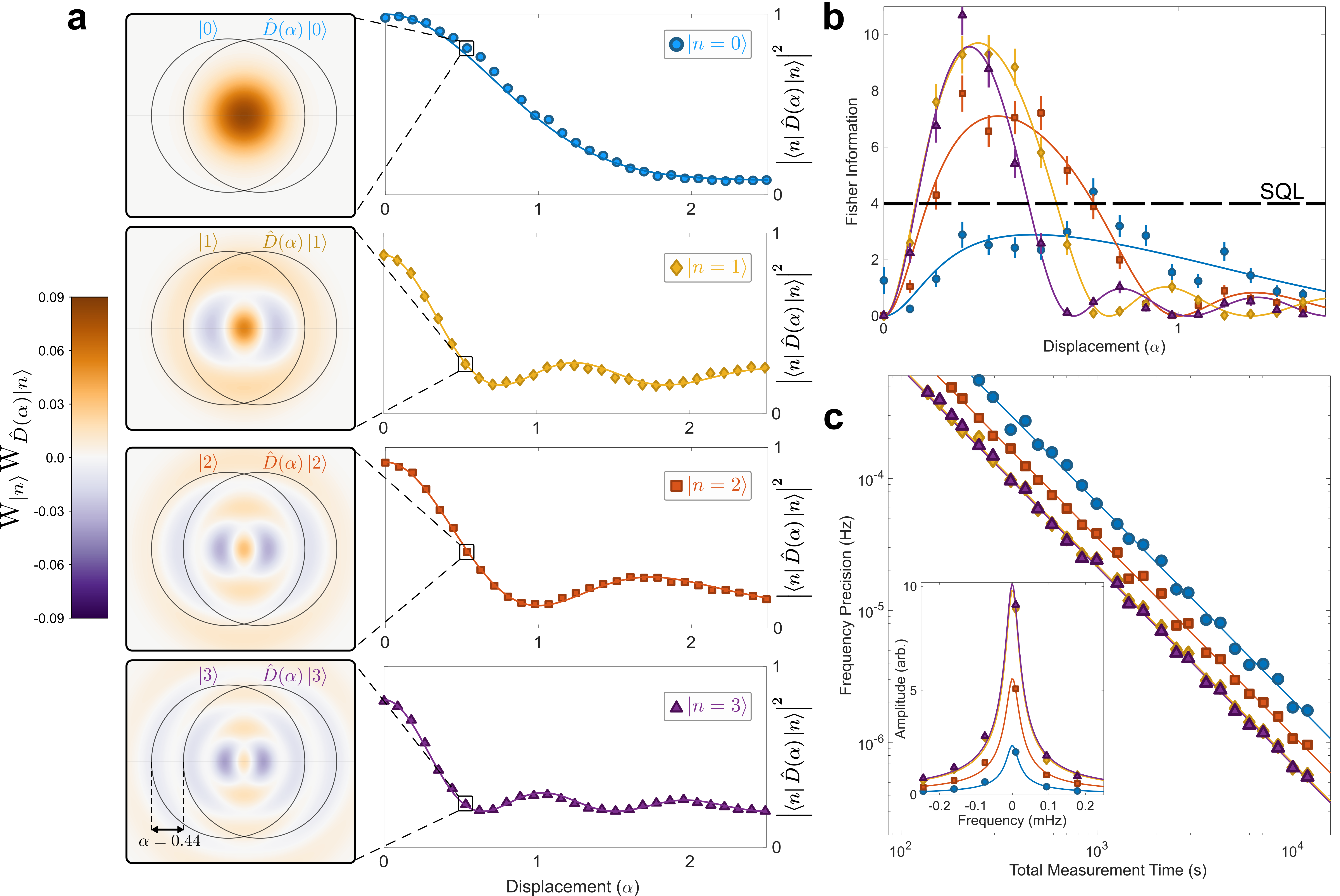}
    \caption{
    \textbf{Quantum-enhanced Qdyne extends metrological gain to arbitrarily long measurement times.}
    Colors denote measurements on the $n=0$ (blue circles), $n=1$ (yellow diamonds), $n=2$ (orange squares), and $n=3$ (purple triangles) Fock states.
    \textbf{(a)} The overlap response (\cref{eq:fock_overlap_prob}) sharpens with $n$ and becomes non-injective as Wigner negativity allows overlap to vanish before states fully separate~\cite{Wolf2019}.
    Solid curves are fits to~\cref{eq:fock_overlap_prob} with additional factors to account for protocol infidelities (\Cref{SI:sec_conventional_spectroscopy}).
    Insets show the product $W_{\ket{n}} W_{\hat{D}(\alpha)\ket{n}}$ at $\alpha = 0.44$, which is integrated to yield the overlap; 90\% contours of the $\ket{n=0}$ Wigner function (solid, black) are shown for scale.
    Points represent 5000 averages; $1\sigma$ error bars are smaller than markers.
    \textbf{(b)} Fisher information extracted from data (points) and fits (solid curves) in (a).
    Departures from the Quantum Fisher Information (QFI) $\mathfrak{F}_n = 4(2n+1)$ grow with $n$ as SPAM errors compound over the more complex preparation sequences required for higher $n$ (\Cref{SI:sec_sql_calculation}); the SQL (black line, dashed) represents $\mathfrak{F}_{n=0}$.
    \textbf{(c)} Microhertz frequency precision with Fock states.
    Precisions $\sigma_\delta/(2\pi)$ are extracted as $1\sigma$ uncertainties on fitted Lorentzian linecenters from DFTs on contiguous data subsets $[0,T]$.
    All $n>0$ states improve on $n=0$ for $\sigma_{\delta}$ across the full range, reaching a minimum $\sigma_\delta/(2\pi) = \qty{0.5}{\micro \Hz}$ with no indication of saturation.
    Linear fits to log-transformed data (solid lines) agree with the expected $T^{-3/2}$ scaling.
    Fast oscillations are spectral leakage tied to shared DFT bin times.
    Inset: DFT spectra show increasing peak amplitude with $n$ from suppression of quantum noise; solid curves are Lorentzian fits.
    Parameters are listed in~\Cref{tab:SI_3c_scaling_values}.
    \label{fig:3_qa_qdyne}
    }
\end{figure*}

The motional Raman Qdyne protocol is summarized in~\Cref{fig:1_technique_overview}.
In the interaction picture of the oscillator, motional Raman excitation heterodynes the signal to a baseband detuning $\delta$, producing the phase-sensitive displacement of \cref{eq:qvsa_displacement} (\Cref{fig:1_technique_overview}(a)).
Tones are applied continuously over an interrogation time $\tau$ (i.e. a Rabi protocol); for detunings $\delta \ll 1/\tau$, the displacement is (up to a global phase)
\begin{equation}    \label{eq:qvsa_displacement_casr}
    \alpha(t) \approx \Omega_q \Omega_d \, \frac{\omega_0 \tau }{\omega_d^2 - \omega_0^2} \,
    \cos{\!\left(\phi_d(t) - \phi_+\right)},
\end{equation}
which imprints the instantaneous signal phase $\phi_d(t) = \phi_d(0) + \delta t$ onto each measurement, where $t$ is the time at which the measurement is performed (\Cref{fig:1_technique_overview}(b, top)).
Phase-tracking of the quadrupole tones (via free evolution of phase accumulators) ensures successive displacements $\alpha_i$ at times $t_i$ remain mutually coherent for reconstruction via Fourier analysis.
The motional state is read out by projection onto the $n$th Fock state (\Cref{fig:1_technique_overview}(b, bottom)), yielding a binary outcome $m_i$ with dark-state probability (\Cref{SI:sec_methods})
\begin{equation}    \label{eq:fock_overlap_prob}
    P_n(\alpha) = \lvert \bra{n} \hat{D}(\alpha) \ket{n} \rvert^2 = e^{-\lvert \alpha \rvert^2} \, L_n\!\left(\lvert \alpha \rvert^2\right)^2,
\end{equation}
where $L_n$ is the n\textsuperscript{th} Laguerre polynomial~\cite{Wolf2019}.
Readout thus depends on $\alpha$ only through the phonon number $\lvert\alpha\rvert^2$ added to $\ket{n}$ by the displacement; $\alpha$ is never measured directly.
As the Laguerre polynomials are non-injective, unambiguous readout requires $\lvert \alpha \rvert$ below the first zero of $L_n$ -- for $n = 3$, the highest Fock state used in this work, this bounds the usable displacement to $\lvert \alpha_{\max} \rvert \lesssim 0.64$.
This does not restrict sensitivity as the Fisher-information optimum lies within this range for all demonstrated $n$ (\Cref{fig:3_qa_qdyne}(b)).
Our use of motional Raman excitation enables this bound to be satisfied for any signal amplitude, as the interaction strength is tunable via the quadrupole tones (\cref{eq:qvsa_displacement}).

We demonstrate Qdyne with motional Raman by reconstructing the spectrum of a dipole signal tone at $\qty{1013}{\MHz}$ heterodyned to a detuning $\delta/(2\pi) = \qty{10}{\Hz}$ (\Cref{fig:1_technique_overview}(c)).
The signal is sampled with period $T_s = \qty{9}{\ms}$ over $T = \qty{900}{\s}$ ($10^{5}$ samples); a discrete Fourier transform (DFT) of the measurement record $\{m_i\}_{1}^{N}$ recovers the signal as a single peak within the Nyquist-limited bandwidth $1/(2 T_s)$.
As readout (\cref{eq:fock_overlap_prob}) depends on $\lvert\alpha\rvert^2$, the reconstructed peak appears at $2\delta/(2\pi) = \qty{20}{\Hz}$ rather than $\delta/(2\pi) = \qty{10}{\Hz}$; the signal frequency is straightforwardly recovered by halving.
This doubling is not fundamental and could be avoided by a readout linear in $\alpha$, e.g.\ the phase-sensitive red sideband protocol~\cite{Hempel2013,Burd2019}.
By symmetry, the spectrum is identical under $\delta \rightarrow -\delta$, though this ambiguity can be resolved by observing peak frequency shifts under small changes $\mu$ to the quadrupole frequencies, i.e. $\omega_d \pm \omega_0 \rightarrow \left(\omega_d + \mu\right) \pm \omega_0$.

\begin{figure}[h!]
	\centering
	\includegraphics[width=\columnwidth]{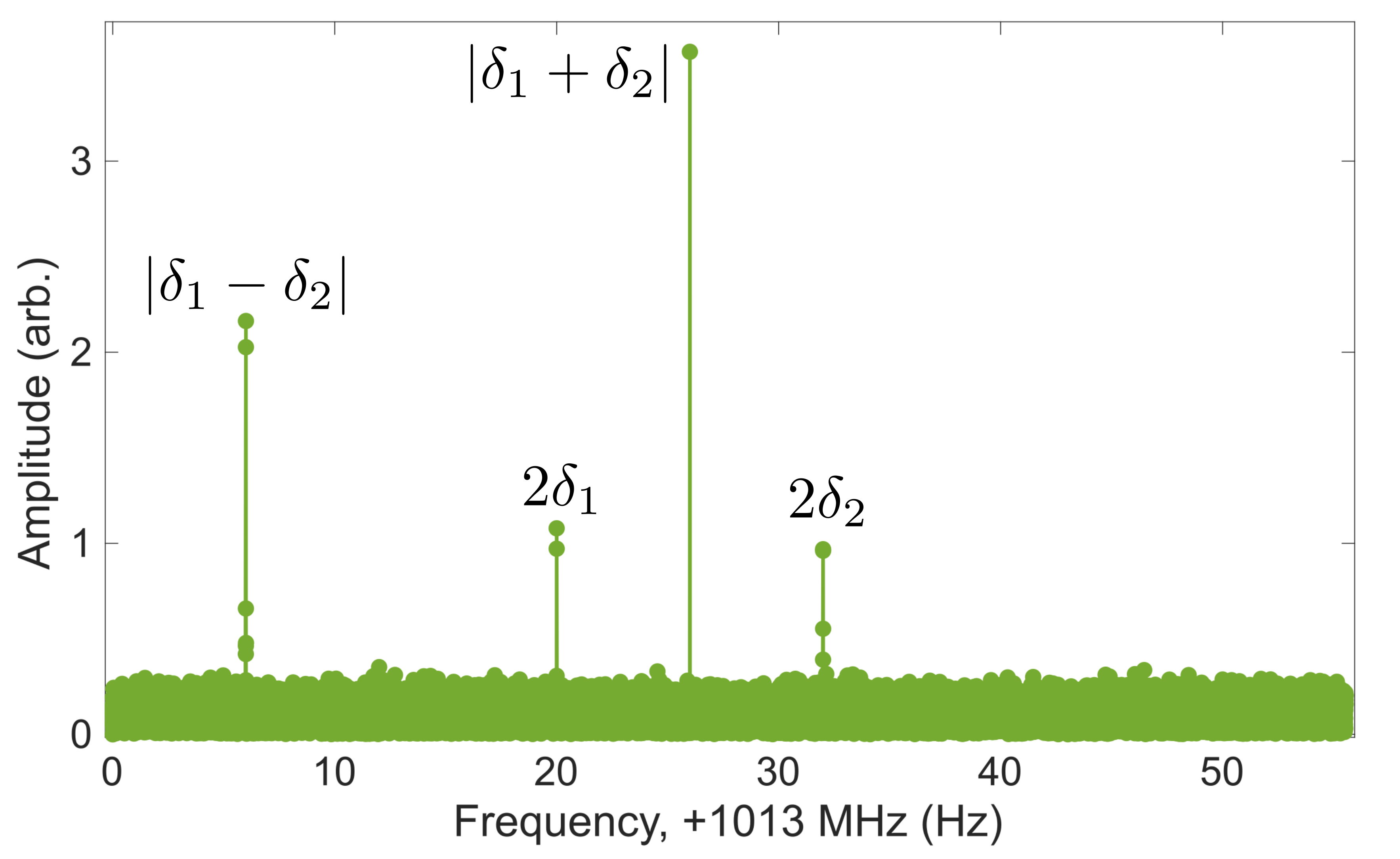}
	\caption{
    \textbf{Simultaneous reconstruction of a multi-tone spectrum.}
    Two equal-amplitude dipole tones at $\omega_d/(2\pi) = \qty{1013}{\MHz}$ are heterodyned to $\delta_{1,2}/(2\pi) = \qtylist{10;16}{\Hz}$ and sampled; acquisition parameters follow~\Cref{fig:1_technique_overview}(c).
    Overlap readout responds to $\left|\alpha(t)\right|^2 = \left|\sum_{i}{a_i \cos{(\delta_i t)}}\right|^2$, which suppresses fundamentals at $\delta_{1,2}$; produces second harmonics at $2\delta_{1,2}/(2\pi) = \qtylist{20;32}{\Hz}$ (amplitude $a_i^2/2$); and intermodulation products at $\left|\delta_1 \pm \delta_2\right|/(2\pi) = \qtylist{6;26}{\Hz}$ (amplitude $a_1 a_2$).
    All peaks exhibit good agreement with predicted frequencies and amplitude ratios; residual asymmetries reflect peak alignments with DFT frequency bins.
    \label{fig:2_multitone}
	}
\end{figure}

Beyond precision, Qdyne enables parallel detection of multiple signals, within the Nyquist $\lvert \delta \rvert/(2 \pi) < 1/(2 T_s)$ bandwidth, without exact knowledge of their frequencies.
\Cref{fig:2_multitone} shows the spectrum of two equal-amplitude dipole tones near $\qty{1013}{\MHz}$ heterodyned to $\delta_{1,2} / (2\pi) = \qtylist{+10;+16}{\Hz}$.
Again, as readout responds to $\left| \alpha(t)\right|^2 = \left|a_1\cos(\delta_1 t) + a_2\cos(\delta_2 t)\right|^2$, the spectrum contains intermodulation products at $\left|\delta_1 \pm \delta_2\right|/(2\pi) = \qtylist{6;26}{\Hz}$ alongside second harmonics at $2\delta_{1,2}/(2\pi) = \qtylist{20;32}{\Hz}$.
Dense, band-limited spectra can nonetheless be recovered via turnpike reconstruction~\cite{TurnpikeProblemSeminal,TurnpikeAlgorithmConditions2017,TurnpikeAlgorithmsNoisyData2024}, while signals separated beyond the Nyquist bandwidth can be recovered by compressive multirate sampling~\cite{DegenCASR2017,Degen2017}.
Equal amplitudes are chosen for clarity only; for arbitrary amplitudes (within the linear range of~\cref{eq:fock_overlap_prob}), peak frequencies remain unchanged and the spectrum can be recovered identically.

These \qty{1013}{\MHz} signals lie roughly two orders of magnitude in frequency beyond the ${\lesssim}\qty{20}{\MHz}$ ceiling of pulsed dynamical-decoupling protocols, which limit the frequency range to the achievable $\pi$-pulse times~\cite{RetzkerQdyne2017,DegenCASR2017,NanoNMRReviewBucher2022,NarrowBandwidthCDD2017Retzker}.
Quantum frequency downconversion with motional Raman thus opens the GHz band to Qdyne without prohibitively fast, high-power probe pulses~\cite{HighFreqQdyne2021Meinel,NanoNMRReviewBucher2022,NarrowBandwidthCDD2017Retzker}.
While a similar wideband extension of Qdyne using quantum frequency mixing~\cite{QFMGuoqing2022} was recently demonstrated on NV ensembles up to \qty{4}{\GHz}~\cite{QFMCASRWalsworth2025}, that demonstration relied on classical ensemble averaging, whereas this work requires only a single bosonic mode and remains available where ensemble averaging is not, e.g. single-ion quantum logic spectroscopy.

Wideband operation alone, however, does not improve \textit{ultimate} sensitivity, which requires use of nonclassical states.
To this end, we integrate Qdyne with Fock-state enhancement for the $n = 0$ to $n = 3$ states.
Fock states are chosen for their rotational symmetry in phase space, which guarantees equal enhancement at all signal phases; rotationally asymmetric states such as cat states~\cite{Hempel2013,Milne2019} would imprint phase-dependent gain and distort reconstruction of dense spectra~\cite{WalsworthCASR2018}.
Enhancement is benchmarked in \Cref{fig:3_qa_qdyne}(a) by measuring the overlap $P_n$ (\cref{eq:fock_overlap_prob})~as a function of displacement $\alpha$.
The overlap response sharpens with increasing $n$ and becomes non-injective due to Fock-state Wigner negativity, which allows the overlap to vanish before the states fully separate~\cite{Wolf2019}.
Metrological gain is quantified in~\Cref{fig:3_qa_qdyne}(b) by computing from the data the Fisher information 
$F{\left(\alpha\right)} = \sum_{i} P{(m_i \rvert \alpha)} \left(\partial_\alpha\!\ln\!{P{(m_i \rvert \alpha)}}\right)^{2}$,
where $m_i$ are measurement outcomes.
The ideal gain is set by the quantum Fisher information (QFI) of the Fock state, $\mathfrak{F}_{n} = 4\left(2n+1\right)$~\cite{Wolf2019,MetrologyBosonic2025Review}, with the SQL defined as $\mathfrak{F}_{n=0}$ (dashed black line,~\Cref{fig:3_qa_qdyne}(b)); departures from the QFI grow with $n$ as state-preparation-and-measurement (SPAM) errors compound over the more complex preparation sequences.
This gain does not reflect the added Fock-state energy, as $\alpha$ is set by the signal and cannot be exchanged for Fock-state quanta; moreover, even at fixed total energy $E = n + \lvert\alpha\rvert^2$, $F_n \propto (2n+1)\lvert\alpha\rvert^2 > F_{n=0}$ for $0 < n < E - \tfrac{1}{2}$.

\Cref{fig:3_qa_qdyne}(c) shows the frequency precision $\sigma_{\delta}/(2\pi)$ as a function of $T$, extracted as the $1\sigma$ uncertainty on linecenters from Lorentzian-type fits to amplitude DFT spectra of contiguous data subsets $[0,T]$; fitting the amplitude rather than the power spectrum yields a linecenter uncertainty independent of the fit window (\Cref{SI:sec_stats_nllsq}).
For all $n>0$, $\sigma_{\delta}$ is improved over the $n=0$ case across the full range of $T$ and is free of the dephasing limits that bound conventional quantum-enhanced spectroscopy at long interrogation times (\Cref{fig:SI_conventional}(b)).
Minimum precisions $\sigma_{\delta}/(2\pi) = \qtylist{1.7;1.0;0.6;0.5}{\micro\Hz}$ are reached with the $n = 0,1,2,3$ states, respectively, at $T = \qty{1.18e4}{\s}$.
Spectral resolution (\Cref{fig:3_qa_qdyne}(c), inset) remains Fourier-limited for all Fock states as Fock-state enhancement suppresses quantum noise to improve SNR, rather than accelerating underlying dynamics.

Frequency precisions in~\Cref{fig:3_qa_qdyne}(c) scale as $T^{-3/2}$, confirmed by unconstrained linear fits to the log-transformed data (\Cref{tab:SI_3c_scaling_values}), in agreement with prediction (\Cref{SI:sec_sql_calculation}) and previous works~\cite{RetzkerQdyne2017,DegenCASR2017}.
Metrological gains are extracted from intercepts of fits to a $T^{-3/2}$ model (\Cref{tab:SI_3c_scaling_values}) and, following calibration, give gains over the $n=0$ state of \qtylist{4.7(1.2);6.2(1.0);7.1(1.0)}{\dB} for the $n=1,2,3$ states, in agreement with Fisher information predictions (\Cref{SI:sec_gain_disparity,SI:sec_gain_estimation}).

Precisions show no indication of saturation as signal and probe tones are digitally synthesized from a common source, such that the $\omega_d/(2\pi) = \qty{86}{\MHz}$ carrier is common-mode and the reference clock instability that limited prior work~\cite{DegenCASR2017, RetzkerQdyne2017} is rejected; termination of acquisition at $T = \qty{1.18e4}{\s}$ reflects experimental overhead and is not fundamental.
The recovered $\sigma_{\delta}/(2\pi) = \qty{0.5}{\micro \Hz}$ is thus an absolute precision set by the measurement rather than the clock, and corresponds to a fractional precision $\sigma_{\delta}/\omega_d \sim 6 \times 10^{-15}$ when referenced to the carrier.
This is not an absolute \textit{accuracy}, as attaining the same precision for an independently generated signal would require a reference itself fractionally stable to $\sim 10^{-15}$.

This quantum-enhanced Qdyne could also be implemented on solid-state spin-defect ensemble sensors -- e.g. spin squeezing has recently been demonstrated on NV centers~\cite{JayichNVSpinSqueezing2025}.
Where bandwidth is prioritized, interrogation times can operate below the $T_2$-limited sensitivity optimum, with quantum enhancement helping to recover the forfeited sensitivity; where sensitivity is prioritized, enhancement instead accelerates signal accumulation, shortening interrogation times and sample periods to extend the bandwidth.
The added bandwidth improves the effective resolution, as larger magnetic bias fields can be used to separate spectral features (e.g. chemical shifts in nanoNMR), while shorter interrogation times reduce susceptibility to low-frequency noise~\cite{NanoNMRReviewBucher2022}.
Quantum frequency downconversion further lifts this constraint, enabling detection above \qty{1}{\GHz} without commensurately fast pulses.

This work promises similar improvements to QHO-based sensors.
Quantum-enhanced spectrum analyzer protocols in trapped ions~\cite{Milne2019, KellerQHOSpecanal2021} forfeit resolution to their nonclassical states -- this could be countered by integration of Qdyne, improving the detection and mitigation of noise sources.
Finally, while we have considered only uncorrelated Markovian dephasing, other noise channels (e.g. motional heating) offer different tradeoffs between resolution, sensitivity, and bandwidth~\cite{Turchette2000, MetrologyBosonic2025Review}; this can be paired with engineered reservoirs~\cite{Myatt2000,ResEng2015Kienzler} to make quantum-enhanced Qdyne a sensitive probe of decoherence itself.


\section{Acknowledgments}
This work was supported by the AFOSR (130427-5114546), the NSF (PHY-2110421 and OMA-2016245), and the ARO (W911NF-24-1-0379).

\section{Competing Interests}
The authors declare no competing interests.

\bibliography{ref}

\appendix

\setcounter{secnumdepth}{1}                         
\renewcommand{\appendixname}{Supplementary Note}    
\counterwithout{equation}{section}                  
\setcounter{equation}{0}\setcounter{figure}{0}\setcounter{table}{0} 
\renewcommand{\theequation}{S\arabic{equation}}     
\renewcommand{\thefigure}{S\arabic{figure}}
\renewcommand{\thetable}{S\arabic{table}}
\onecolumngrid      
\include{main_SI}   

\end{document}

%% file: main_SI.tex
\title{Supplementary Information}
\maketitle
\onecolumngrid

\vspace{-2em}
\section{Experimental Methods}    \label{SI:sec_methods}
\subsection{Signal Generation} \label{SI:methods_signal_gen}
\begin{figure*}[htp!]
    \centering
    \includegraphics[width=\textwidth]{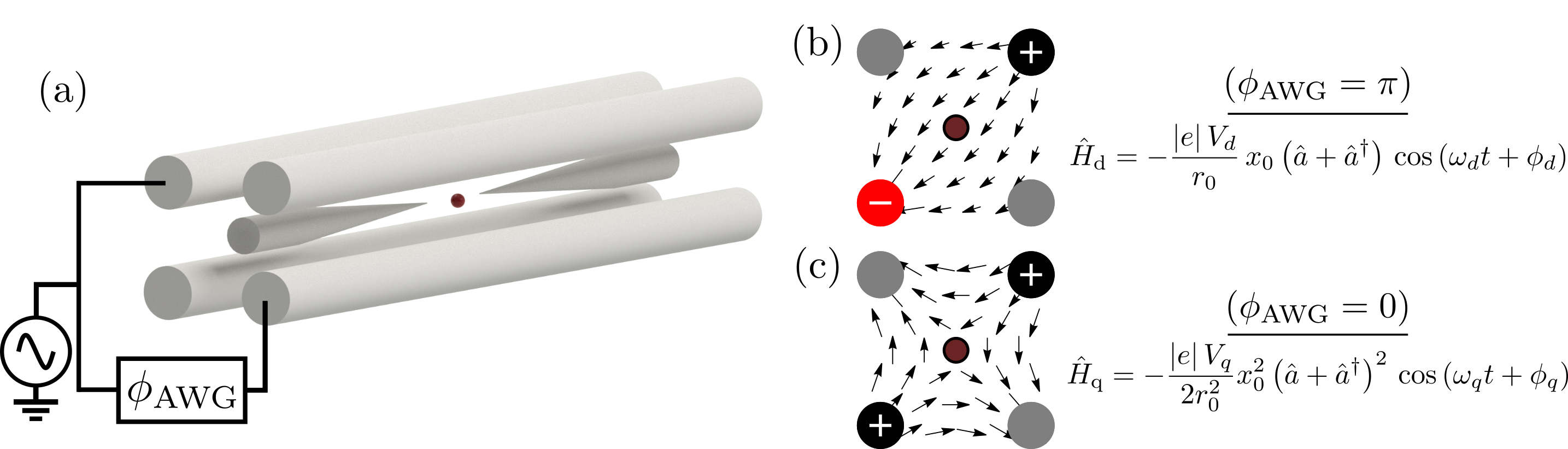}
    \caption{Implementation of motional Raman excitation on the radial modes.
    \textbf{(a)} Oscillating voltages from two synchronously-clocked channels of an arbitrary waveform generator (AWG) are coupled onto diametrically opposed radial electrodes.
    \textbf{(b)} A dipole field is generated for a relative phase $\phi_\textrm{AWG} = \pi$ between the voltages coupled to the electrodes, while a quadrupole field \textbf{(c)} is generated if the coupled voltages are exactly in phase (i.e. $\phi_\textrm{AWG} = 0$).
    \label{fig:SI_schematic}
    }
\end{figure*}

The experiment is controlled using the Advanced Real-Time Infrastructure for Quantum physics (ARTIQ) control system.
Control signals are generated by a Field-Programmable Gate Array board (Kasli v2.0), which synchronously clocks all peripherals to a \qty{125}{\MHz} crystal reference.
Radio-frequency (RF) signals used for motional excitation are generated using the Phaser arbitrary waveform generator (AWG).
Both outputs of the Phaser AWG are passed through a signal conditioning stage before being coupled to two diametrically opposing trap electrodes (\Cref{fig:SI_schematic}).

\subsection{Fock State Enhancement Protocol} \label{SI:methods_fock}
The Fock state enhancement protocol follows~\cite{Wolf2019}: an initial Fock state $\ket{n}$ is prepared, a displacement $\hat{D}(\alpha)$ is applied, and the overlap with the same state $\ket{n}$ is measured.

Fock states $\ket{n}$ are prepared by alternately exciting the blue and red motional sidebands of the narrow $\left.^{2}\textrm{S}_{1/2}\right. \leftrightarrow \left.^2\textrm{D}_{5/2}\right.$ qubit transition.
For odd $n$, an additional carrier $\pi$-pulse follows the final sideband pulse to return the spin to the $\left.^{2} \textrm{S}_{1/2} \right.$ manifold and suppress spontaneous emission during the sequence.

\begin{figure*}[htp!]
    \centering
    \includegraphics[width=\textwidth]{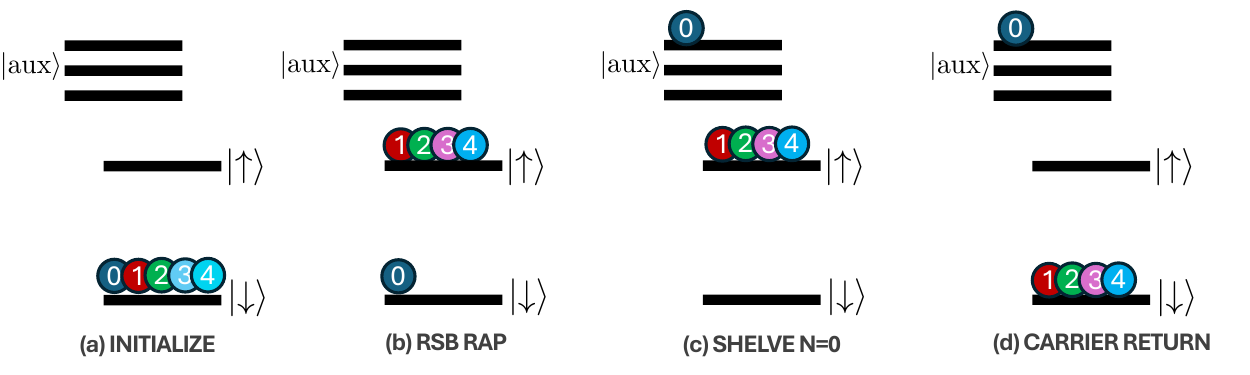}
    \caption{Fock state overlap readout protocol.
    A series of sideband Rapid Adiabatic Passage and shelving pulses isolate a single Fock state population in $\ket{\downarrow}$ for state detection; auxiliary states used are the $m_J = -3/2, +1/2, +3/2$ sublevels of the $\left.^{2}\textrm{D}_{5/2}\right.$ manifold.
    \textbf{(a)} The target motional state distribution is prepared.
    \textbf{(b)} A RAP pulse on the RSB transfers $\ket{\downarrow, n} \rightarrow \ket{\uparrow, n-1}$, leaving $p_0$ in $\ket{\downarrow}$ -- the protocol can be terminated here to detect the $p_0$ population (via e.g. state-selective fluorescence), or continued to detect higher $p_n$.
    \textbf{(c)} The $p_0$ population is shelved into an auxiliary $\left.^{2}\textrm{D}_{5/2}\right.$ Zeeman sublevel via carrier $\pi$-pulse
    to remove it from subsequent dynamics.
    \textbf{(d)} A carrier $\pi$-pulse returns $\ket{\uparrow, n-1} \rightarrow \ket{\downarrow, n-1}$ for the cycle to repeat from (b) until the target $p_n$ is isolated.
    \label{fig:SI_fock_overlap}
    }
\end{figure*}

Motional overlap readout is shown in~\Cref{fig:SI_fock_overlap} and is implemented by isolating the $n$-th motional state in the $\textrm{S}_{1/2}$ manifold and reading it out via state-selective fluorescence.
For a given motional state distribution, a Rapid Adiabatic Passage (RAP) pulse on the red motional sideband~\cite{PhononArithmetic2016Kihwan} transfers all population except that in $\ket{n = 0}$ to the $\textrm{D}_{5/2}$ manifold, isolating it in the $\textrm{S}_{1/2}$ manifold.
This population is shelved in an auxiliary state, after which the remaining $\textrm{D}_{5/2}$ population is returned to $\textrm{S}_{1/2}$.
This sequence is repeated until the target Fock state population is isolated in the $\textrm{S}_{1/2}$ state, at which point state-selective fluorescence reads out the overlap.
To improve robustness to pulse area and detuning errors and to reduce calibration overhead, all sideband, carrier, and shelving pulses are implemented as RAP pulses, achieving individual pulse fidelities of approximately $97\%$.
Use of RAP requires an overhead $\sim\qty{1}{\ms}$ per Fock state, accounting for $\approx 30\%$ of the total enhancement overhead.
Shelving uses the $m_J = -3/2, +1/2, +3/2$ Zeeman sublevels of the $\textrm{D}_{5/2}$ manifold; the highest accessible Fock state is limited by the number of available auxiliary states, which in our implementation is thus $n \leq 3$.
This could be substantially extended by adopting improved number-resolving motional-readout techniques, such as ultranarrowband pulse sequences~\cite{UNBMotionalReadout2024Blatt}.

\subsection{Uncertainty Analysis for the Fisher Information in~\Cref{fig:3_qa_qdyne}(b)} \label{SI:ssec_error_fisher_3b}
Uncertainties for the Fisher information curves in~\Cref{fig:3_qa_qdyne}(b) are analyzed following~\cite{Wolf2019}.
Fisher information values in~\Cref{fig:3_qa_qdyne}(b) are extracted from~\Cref{fig:3_qa_qdyne}(a) using symmetric difference quotients, and asymmetric difference quotients for the endpoints:
\begin{align}   \label{eq:SI_error_slope_eqn}
    s_i = \frac{P_{i+1} - P_{i-1}}{\alpha_{i+1} - \alpha_{i-1}}, \qquad
    s_i = \frac{P_{i+1} - P_{i}}{\alpha_{i+1} - \alpha_{i}}.
\end{align}
The Fisher information $F_{\alpha} = \left(\partial_{\alpha}P\right)^2/\sigma_P^2$ is then
\begin{equation}
    F_{i} = \frac{s_i^2}{(\sigma_P^2)_i} = \frac{s_i^2}{P_i \left(1 - P_i\right)},
\end{equation}
with the uncertainty obtained by standard error propagation
\begin{align}
    \left(\Delta{F}\right)^2 &=
        \underbrace{\left(\frac{\partial F}{\partial s} \Delta s_{\textrm{Total}} \right)^2}_{\textrm{Slope}} +
        \underbrace{\left(\frac{\partial F}{\partial \sigma_P^2} \Delta \sigma_P^2 \right)^2}_{\textrm{QPN}} \nonumber \\
    &= \left( \frac{2s}{\sigma_P^2} \Delta s_{\textrm{Total}}\right)^2 + \left(\frac{s^2}{\sigma_P^4} \Delta \sigma_P^2\right)^2 \nonumber \\
    &= F^2 \left(4 \left(\frac{\Delta s_{\textrm{Total}}}{s}\right)^2 + \left(\frac{\Delta\sigma_P^2}{\sigma_P^2}\right)^2\right)
\end{align}
where $\Delta s_{\textrm{Total}}$ is the total uncertainty from slope determination, and $\Delta \sigma_P^2$ is the uncertainty of the variance from Quantum Projection Noise (QPN).

\subsubsection{Uncertainties in Slope Determination}
From~\cref{eq:SI_error_slope_eqn}, the total uncertainty on the slope $\Delta s_{\textrm{Total}}$ is
\begin{align}    \label{eq:SI_error_slope_full}
    \left(\Delta s_{\textrm{Total}}\right)^2 &=
    \underbrace{
        \left(\frac{\partial s}{\partial P_{i+1}} \Delta P_{i+1}\right)^2 +
        \left(\frac{\partial s}{\partial P_{i-1}} \Delta P_{i-1}\right)^2
    }_{\left(\Delta s\right)^2_{\textrm{QPN}}} +
    \underbrace{
        \left(\frac{\partial s}{\partial \alpha} \Delta \alpha\right)^2
    }_{\left(\Delta s\right)^2_{\alpha}} \nonumber \\
    &= \left(\frac{1}{\alpha_{i+1} - \alpha_{i-1}} \Delta P_{i+1}\right)^2 +
        \left(\frac{1}{\alpha_{i+1} - \alpha_{i-1}} \Delta P_{i-1}\right)^2 +
        \left(\frac{s}{\left(\alpha_{i+1} - \alpha_{i-1}\right)} \Delta\alpha\right)^2 \nonumber \\
    &= s^2 \left(
        \frac{\Delta P_{i+1}^2 + \Delta P_{i-1}^2}{\left(P_{i+1} - P_{i-1}\right)^2} +
        \left(\frac{\Delta \alpha}{\alpha_{i+1} - \alpha_{i-1}}\right)^2
    \right)
\end{align}
where $\Delta P_i$ is the uncertainty from Quantum Projection Noise (QPN).

\subparagraph{Quantum Projection Noise}
Values for $\Delta P_i$ are determined using the Wilson interval to improve estimation near extrema (i.e. $P=0,1$), though we recenter the interval to simplify error propagation, i.e.
\begin{equation}
    \Delta P_i = \frac{\sqrt{\frac{1}{4 N^2} + \frac{P_i (1-P_i)}{N}}}{\frac{1}{N} + 1}.
\end{equation}

\subparagraph{Uncertainty in $\alpha$}
Interrogation times are converted to displacements by a scale factor $\alpha = \dot{\alpha} \tau$, with $\dot{\alpha}$ determined by a fit to the data with~\cref{eq:fock_prob_fit}.
The uncertainty in $\alpha$ is therefore
\begin{equation}
    (\Delta \alpha)^2 = \left( \frac{\alpha_{i+1} - \alpha_{i}}{\dot{\alpha}} \Delta \dot{\alpha} \right)^2,
\end{equation}
where $\Delta\dot{\alpha}$ is the uncertainty in the fitted $\dot{\alpha}$.

\subparagraph{Finite Difference Quotient}
The finite difference quotient contributes an error
\begin{equation}
    e = \frac{f^{(3)}{(x)}}{3!} h^2
\end{equation}
where $h$ is the step size.
As this error is systematic, it is subtracted from $\hat{s}$ and does not enter~\cref{eq:SI_error_slope_full} (cf.~\cite{Wolf2019}), i.e.
\begin{equation}
    \hat{s}' = \hat{s} - \frac{P^{(3)}{(\alpha_i)}}{3!} \left(\frac{\alpha_{i+1} - \alpha_{i-1}}{2}\right)^2,
\end{equation}
where we have used $h = (\alpha_{i+1} - \alpha_{i-1})/2$, and $P$ is obtained from the fitted function (\cref{eq:fock_prob_fit}).

\subsubsection{Uncertainty in the QPN Variance}
The variance of the sample variance for $N$ random iid samples $\left(X_1, \ldots, X_N\right)$ with arbitrary distribution is
\begin{equation}
    \textrm{Var}{\left(S^2\right)} = \frac{1}{N} \left(\theta_4 - \frac{N-3}{N-1}\theta_2^2\right)
\end{equation}
where $\theta_k = \mathbb{E}\left[(X - \mu)^k\right]$ is the $k$th central moment~\cite{StatsCasellaBerger}.
If the $X_i$ are Bernoulli distributed, this evaluates to
\begin{equation}
    \left(\Delta\sigma_P^2\right)^2 \equiv \textrm{Var}{\left(\sigma_P^2\right)} = \frac{\sigma_P^2}{N} \left[1 - \sigma_P^2 \left(4-\frac{2}{N-1}\right)\right] \approx \frac{\sigma_P^2 \left( 1-4\sigma_P^2\right)}{N}.
\end{equation}

\clearpage
\subsection{Figure Parameters} \label{SI:ssec_figure_parameters}
Parameters and fit results for the figures in the main text are given here.

\begin{table*}[hp!]
    \centering
    \begin{tabular}{c c c}
    (a) & \hspace{3mm} & (b) \\
    {\begin{tabular}{|c|c|c|c|c|}
        \hline
         & $\alpha_{0}$ & $F_{n}$ & $g_{n=0}$ (dB) & $g_{\textrm{SQL}}$ (dB) \\
        \hline
        $\ket{n=0}$ & 0.711 & $4.4(5)$ & N/A & $0.4(5)$ \\
        $\ket{n=1}$ & 0.267 & $7.9(6)$ & $2.5(6)$ & $3.0(4)$ \\
        $\ket{n=2}$ & 0.356 & $9.3(7)$ & $3.2(6)$ & $3.7(3)$ \\
        $\ket{n=3}$ & 0.267 & $10.7(7)$ & $3.8(5)$ & $4.3(3)$ \\
        \hline
    \end{tabular}}
    & &
    {\begin{tabular}{|c|c|c|c|}
        \hline
         & $F_{n}$ & $g_{n=0}$ (dB) & $g_{\textrm{SQL}}$ (dB) \\
        \hline
        $\ket{n=0}$ & $2.4(4)$ & N/A & $-2.2(7)$ \\
        $\ket{n=1}$ & $7.0(6)$ & $4.6(8)$ & $2.5(4)$ \\
        $\ket{n=2}$ & $8.8(7)$ & $5.6(7)$ & $3.5(3)$ \\
        $\ket{n=3}$ & $5.4(5)$ & $3.5(8)$ & $1.3(4)$ \\
        \hline
    \end{tabular}}
    \end{tabular}
    \caption{Quantum-enhanced gains extracted from data in~\Cref{fig:3_qa_qdyne}(b): \textbf{(a)} maximized over $\alpha$ and \textbf{(b)} at $\alpha=0.44$.
    $g_{n=0}$ represents the gain over the $n=0$ state and $g_{\textrm{SQL}}$ represents the gain over the SQL $\mathfrak{F}_{n=0}=4$ (\cref{eq:SI_sql_QFI_higherFock}).
    Gains are obtained as the Fisher information ratio $10\log_{10}{\left(\frac{F_{m}}{F_n}\right)}$.
    \label{tab:SI_3b_fisher_vals}
    }
\end{table*}

\begin{table*}[hp!]
    \centering
    \begin{tabular}{|c|c|c|c|c|c|c|c|c|c|}
        \hline
         & $\dot{\alpha}$ & $A_0$ & $A_1$ & $A_2$ & $A_3$ &
            $C_0$ & $C_1$ & $C_2$ & $C_3$ \\
        \hline
        \Cref{fig:3_qa_qdyne}(a) & $2.669(13)$ & $0.919(4)$ & $0.795(8)$ & $0.719(9)$ & $0.618(10)$ &
            $0.001(3)$ & $0.081(6)$ & $0.122(8)$ & $0.183(9)$ \\
        \Cref{fig:3_qa_qdyne}(b) & --- & $0.945(19)$ & $0.803(8)$ & $0.726(3)$ & $0.611(6)$ &
            $0.040(13)$ & $0.082(10)$ & $0.127(6)$ & $0.218(15)$ \\
        \hline
    \end{tabular}
    \caption{Experimental parameters extracted by fitting models of the form~\Cref{eq:fock_prob_fit} to~\Cref{fig:3_qa_qdyne}(a,b).
    A single value of $\dot{\alpha}$ is fitted jointly across all Fock states as it is fixed by hardware timing and does not vary between runs.
    \label{tab:SI_ADJCDJ_fig3a}}
\end{table*}

\begin{table*}[hp!]
    \centering
    \begin{tabular}{c c c c c}
    (a) & \hspace{3mm} & (b) & \hspace{3mm} & (c)\\
    {\begin{tabular}{|l|l|}
        \hline
        Parameter & Value\\
        \hline
        Number of Samples & 890582 \\
        Dipole Carrier Freq. & \qty{86}{\MHz} \\
        Heterodyned Beat Freq. & \qty{10}{\Hz} \\
        Sample Period & \qty{13.3}{\ms} \\
        Interrogation Time & \qty{300}{\us} \\
        Displacement ($n=0$) & 0.32(2) \\
        Displacement ($n=1$) & 0.31(2) \\
        Displacement ($n=2$) & 0.40(2) \\
        Displacement ($n=3$) & 0.38(2) \\
        \hline
    \end{tabular}}
    & &
    {\begin{tabular}{|c|| r|r || r |}
        \hline
        & \multicolumn{2}{|c||}{\textbf{Free Fit}} &
            \multicolumn{1}{|c|}{\textbf{Fixed Fit}} \\
        \hline
         & $\log b$ & $m$ & $\log b$ \\
        \hline
        $\ket{n=0}$ & 0.55(7) & -1.486(9) & 0.659(12) \\
        $\ket{n=1}$ & 0.25(3) & -1.514(4) & 0.136(6) \\
        $\ket{n=2}$ & -0.33(4) & -1.504(5) & -0.366(6) \\
        $\ket{n=3}$ & -0.37(3) & -1.504(3) & -0.397(4) \\
        \hline
    \end{tabular}}
    & &
    {\begin{tabular}{|c|| r|r |}
        \hline
        & \multicolumn{2}{|c|}{\textbf{Free Fit}} \\
        \hline
         & $\log b$ & $m$ \\
        \hline
        $\ket{n=0}$ & -1.62(7) & -0.980(9) \\
        $\ket{n=1}$ & -1.64(3) & -0.982(4) \\
        $\ket{n=2}$ & -1.57(4) & -0.988(5) \\
        $\ket{n=3}$ & -1.51(3) & -0.997(4) \\
        \hline
    \end{tabular}}
    \end{tabular}
    \caption{
    Parameters (a) and fit results (b,c) for~\Cref{fig:3_qa_qdyne}(c).
    \textbf{(b)} Frequency precision $\sigma_\delta$ and \textbf{(c)} fitted linewidth $\Gamma$ versus total measurement time $T$, fitted as $\log y = \log b + m \log T$ with $m$ free (left columns) and with $m = -3/2$ fixed (right column, (b)); uncertainties are $1\sigma$.
    Metrological gains are extracted from fixed-slope intercepts (right column, (b)) (see~\Cref{SI:sec_gain_disparity}).
    The displacement uncertainty $\delta\alpha_0 = 0.02$ bounds the observed long-term drift.
    \label{tab:SI_3c_scaling_values}
    }
\end{table*}

\clearpage
\section{The Standard Quantum Limit for Quantum-Enhanced Qdyne} \label{SI:sec_sql_calculation}
\subsection{Fisher Information -- Displacement Sensing}
For the motional overlap readout protocol, the probability of a dark-state measurement given displacement $\alpha$ using the $n=0$ state is
\begin{equation}
    P_{n=0}{(\alpha)} = e^{- \lvert \alpha \rvert^2}.
\end{equation}
The Fisher information is calculated following
\begin{equation}    \label{eq:fisher_info}
    F{\left(\alpha\right)} = \sum_{\mu} \frac{1}{P(\mu \rvert \alpha)} \left(\frac{\partial P(\mu \rvert \alpha)}{\partial \alpha}\right)^{2},
\end{equation}
which leads to
\begin{align}   \label{eq:SI_sql_displ_1}
    F{(\alpha)} &= \frac{1}{\sigma_P^2} \left( \frac{\partial P}{\partial \alpha} \right)^2 \nonumber \\
    &= \frac{4 \lvert \alpha \rvert^2 e^{-2\lvert \alpha \rvert^2}}{e^{-\lvert \alpha \rvert^2} \left( 1-e^{-\lvert \alpha \rvert^2} \right)} 
    = \frac{4 \lvert \alpha \rvert^2}{e^{\lvert \alpha \rvert^2} - 1} \nonumber \\
    &\leq 4.
\end{align}
Our protocol is therefore capable of saturating the Quantum Fisher Information (QFI) bound $\mathfrak{F}=4$~\cite{Wolf2019}.
For higher Fock states, the QFI is~\cite{Wolf2019}
\begin{equation}    \label{eq:SI_sql_QFI_higherFock}
    \mathfrak{F}_n = 4 \, (2n+1).
\end{equation}

\subsection{Fisher Information --- Frequency Sensing (with Displacements)}
We are interested in sensing the frequency of a signal from a displacement
\begin{equation}    \label{eq:SI_sql_displaceSpectrum}
    \alpha(\delta) = \Omega \tau \, \textrm{sinc}{\left(\frac{\delta \tau}{2}\right)}.
\end{equation}
The frequency Fisher information is then
\begin{align}    \label{eq:SI_sql_displace_FI_1}
    F{(\delta)} &= \frac{1}{\sigma_P^2} \left( \frac{\partial P}{\partial \delta} \right)^2 \\
    &= \frac{1}{\sigma_P^2} \left( \frac{\partial P}{\partial \alpha} \frac{\partial \alpha}{\partial \delta} \right)^2
    = \frac{1}{\sigma_P^2} \left( \frac{\partial P}{\partial \alpha} \right)^2  \left(\frac{\partial \alpha}{\partial \delta}\right)^2 \nonumber \\
    &= F(\alpha) \left(\frac{\partial \alpha}{\partial \delta}\right)^2
\end{align}
The term $(\partial \alpha / \partial \delta)$ can be understood as the slope of the spectrum, and is expanded
\begin{equation}    \label{eq:SI_sql_displaceSpectrum_opt}
    \frac{\partial \alpha}{\partial \delta} = \frac{\Omega \tau^2}{2} \, \left( \frac{\cos{(x)} - \textrm{sinc}{(x)}}{x} \right)
\end{equation}
where we have used the convenient substitution $x \equiv \delta \tau /2$.
\Cref{eq:SI_sql_displaceSpectrum_opt} is maximized at $x = 2.08158$, giving the Standard Quantum Limit (SQL) for frequency sensing
\begin{equation}    \label{eq:SI_sql_displace_FI_2}
    F_{\max}(\delta) = 4 \, \left(\frac{k \Omega \tau^2}{2}\right)^2
\end{equation}
where $k = 0.4362$ (from maximization of~\cref{eq:SI_sql_displaceSpectrum_opt}).

\subsection{Fisher Information --- Quantum-Enhanced Qdyne}
Qdyne comprises a sequence of measurements $\{P_0, P_1, \dots , P_N \}$ at times $\{t_0, t_1, \dots, t_N\}$ where $t_j = t_0 + j\,T_s$.
As Fisher information is additive, the collective Fisher information of Qdyne is
\begin{equation}   \label{eq:SI_sql_qdyne_base}
    F_\textrm{tot} = \sum_{j=0}^{N}{F_j(\delta)},
\end{equation}
where $F_j(\delta)$ is the frequency Fisher information of the $j$-th measurement.

The phase-sensitive displacement required for Qdyne can be described as 
\begin{equation}   \label{eq:SI_sql_qdyne_displacement}
    \alpha_j = \alpha_0 \sin{\left( \phi_0 + \delta t_j \right)},
\end{equation}
where $\alpha_0 =  \Omega \tau \, \textrm{sinc}{\left(\frac{\delta\tau}{2}\right)}  \approx \Omega \tau$ in the $\delta\tau \ll 1$ regime of Qdyne, $\phi_0$ is some arbitrary initial phase, and $t_j$ is the time of the $j$-th measurement.
The Fisher information becomes
\begin{align}   \label{eq:SI_sql_qdyne_1}
     F_\textrm{tot} &= \sum_{j=0}^{N}{F(\alpha_j) \left(\frac{\partial \alpha_j}{\partial \delta}\right)^2} \nonumber \\
     &= \sum_{j=0}^{N}{F(\alpha_j) \left( \alpha_0 t_j \cos{(\phi_0 + \delta t_j)} \right)^2},
\end{align}
where the SQL is obtained by setting $F(\alpha_j) = \mathfrak{F}=4$.
The sum in~\Cref{eq:SI_sql_qdyne_1} can be simplified as
\begin{equation}
     t_j^2 \cos^2{(\phi_0 + \delta t_j)}
     \; = \; t_j^2 \left( \frac{1 + \cos{\left(2 (\phi_0 + \delta t_j) \right)}}{2} \right)
     \; \approx \; \frac{t_j^2}{2} \nonumber \\
\end{equation}
where the cosine term averages to zero over the sum.

\Cref{eq:SI_sql_qdyne_1} can then be evaluated as a Riemann sum
\begin{equation}    \label{eq:SI_sql_qdyne_2}
    \frac{1}{\Delta t} \sum_{t_j = t_0}^{t_N} \frac{t_j^2}{2} \Delta t
    \; = \; \frac{1}{T_s} \int_{t_0}^{t_N}{\frac{t^2}{2} \, dt}
    \; = \; \frac{1}{6} \frac{T^3}{T_s}
\end{equation}
where without loss of generality we have set $t_0 = 0$, and used $T = t_N$.
The SQL for Qdyne is therefore
\begin{equation}    \label{eq:SI_sql_qdyne_final}
     F_\textrm{tot}
     \; =\; \left( \mathfrak{F} \, \alpha_0^2\right) \frac{1}{6} \frac{T^3}{T_s}
     \; =\; \frac{2}{3} \alpha_0^2 \frac{T^3}{T_s}.
\end{equation}
For higher Fock states, this is
\begin{equation}    \label{eq:SI_sql_qdyne_final_higherFock}
     F_n
     = (2n+1) \times F_{\textrm{tot}}
     = (2n+1)\frac{2}{3} \alpha_0^2 \frac{T^3}{T_s}
\end{equation}

\clearpage
\section{Conventional Spectroscopy Under Dephasing}    \label{SI:sec_conventional_spectroscopy}
\begin{figure*}[ht!]
    \centering
    \includegraphics[width=\textwidth]{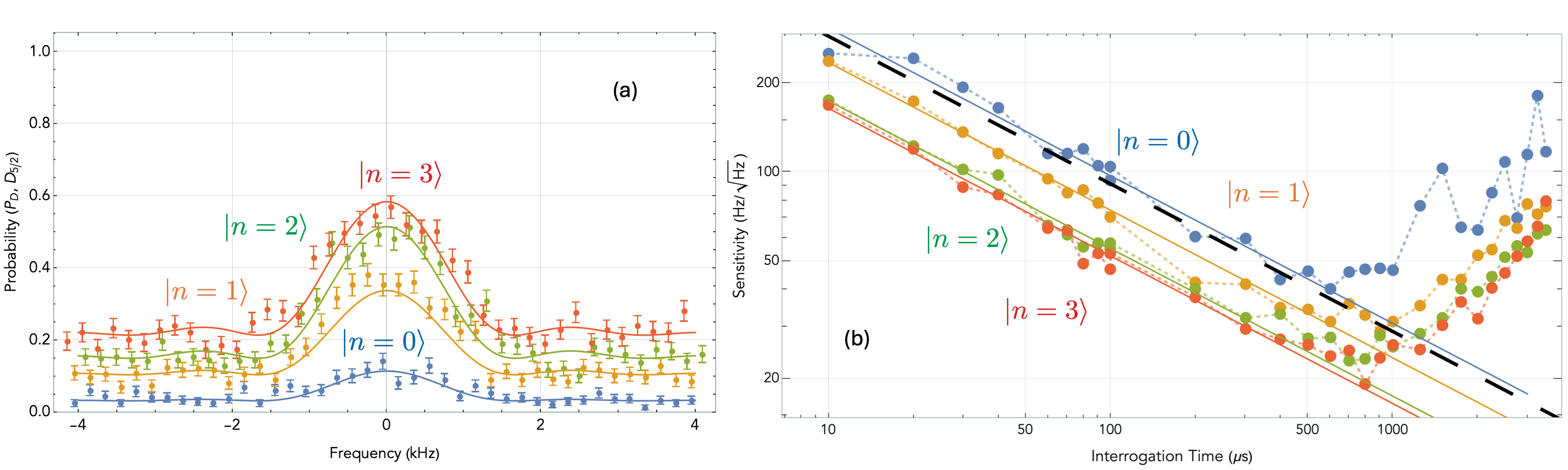}
    \caption{Sensitivity with conventional, quantum-enhanced spectroscopy is dephasing-limited.
    Data for the $n=0,1,2,3$ Fock states are denoted by blue, orange, green, and red, respectively.
    \textbf{(a)} Sample Rabi lineshapes at $\tau = \qty{600}{\us}$ show state overlap $P_n$ as a function of signal detuning $\delta$ for motional Raman excitation at $\omega_d/(2\pi) = \qty{86}{\MHz}$.
    The contrast and background scale with $n$, following~\Cref{fig:3_qa_qdyne}(a); improved frequency sensitivity arises from the improvement in the overall SNR.
    Solid curves are fits to the data using~\cref{eq:fock_prob_fit} with $\alpha=\alpha_0 \, \mathrm{sinc}{\left(\left(\delta - \delta_0\right) \tau/2\right)}$.
    Error bars represent the $1\sigma$ uncertainty; results are averaged over $N=250$ repetitions.
    \textbf{(b)} Frequency sensitivity $\delta_{\min}(\tau) = \sqrt{\tau / F_\delta}$ as a function of interrogation time $\tau$, at fixed displacement $\alpha = 0.4$.
    For each $\tau$, a Rabi lineshape is acquired as in (a) ($N = 200$ repetitions) to extract the frequency Fisher information $F_\delta$.
    Solid lines are linear fits to the log-transformed data for $\tau \le \qty{1000}{\us}$; the dashed black line is the SQL (\cref{eq:SI_sensitivity_conventional_SQL}).
    Above $\tau \gtrsim \qty{1000}{\us}$, dephasing becomes significant and sensitivity deteriorates, limiting the maximum interrogation time.
    \label{fig:SI_conventional}
    }
\end{figure*}

In~\Cref{fig:SI_conventional}, we measure the frequency sensitivity under Fock-state enhancement as a function of interrogation time $\tau$ using quantum-enhanced spectroscopy.
At each $\tau$, a Rabi lineshape is obtained by measuring the state overlap $P_n$ as the detuning is scanned across a range $\delta \sim O(1/\tau)$ with maximum displacement $\alpha_0 \propto \Omega_q\Omega_d \,\tau$ fixed by scaling $\Omega_d,\Omega_q \propto 1/\sqrt{\tau}$ (\cref{eq:qvsa_displacement}).
Lineshapes are smoothed by fitting the analytical form
\begin{equation}    \label{eq:fock_prob_fit}
    P_n = C + A \, S_n, \qquad
    S_n \equiv 1 - e^{-\lvert \alpha \rvert^{2}} L_{n}\!{\left( \lvert \alpha \rvert^{2} \right)}^{2},
\end{equation}
where $\alpha = \alpha_0 \, \textrm{sinc}{\left((\delta - \delta_0 ) \tau/2 \right)}$, with free parameters $\alpha_0$, $\delta_{0}$ and $\tau$.
For each $\tau$, the maximum frequency Fisher information $F_{\delta}{\left(\tau\right)}$ is numerically calculated from the lineshape and scaled by the ideal, zero dead time sample rate $1/\tau$, which yields the frequency sensitivity via the Cram\'{e}r-Rao bound $\delta_{\min} = \sqrt{\tau / F_\delta}$~\cite{Degen2017}.
Though dead time exists experimentally and scales with Fock state, this contribution is discounted as it is an artefact of the enhancement protocol and is not fundamental.
Other nonclassical states require negligible overhead relative to decoherence times -- highly squeezed states of $r=2.54$ can be generated in $\sim \qty{10}{\us}$~\cite{Burd2019}.

Sensitivities for $\alpha_0 = 0.4$ are presented in~\Cref{fig:SI_conventional}(b).
The ideal sensitivity
\begin{equation}    \label{eq:SI_sensitivity_conventional_SQL}
    \delta_{\textrm{min}} = \frac{k}{\alpha_0 \sqrt{\tau}} \frac{1}{\sqrt{4\left(2n+1\right)}},
\end{equation}
where $k=0.73$ is a constant for a Rabi protocol, is plotted for $n=0$ (black, dashed), which further serves as the SQL \cite{Wu2025}.
Sensitivities for all states $\ket{n}$ follow the expected $1/\sqrt{\tau}$ scaling and show good agreement with the analytical form of $\delta_{\textrm{min}}$.
Gains are compared to the $n=0$ state via the intercept of linear models $\log{\left(y\right)} = \log{\left(b\right)} - 0.5\times \log{\left(\tau\right)}$ fitted to the data for $\tau < \qty{1000}{\us}$ (solid lines), and yield $g_{n=1}=\qty{2.3(4)}{\dB} ,\, g_{n=2}=\qty{4.9(5)}{\dB} ,\, g_{n=3}=\qty{5.4(4)}{\dB}$.

In~\Cref{fig:SI_conventional}(b), all Fock states reach optimal sensitivity at similar $\tau \approx \qty{1000}{\us}$ as the accelerated decoherence (specifically, dephasing) that ordinarily penalizes higher-energy states is absent in this small-displacement regime.
This contrasts with GHZ states~\cite{GHZClockDecoherence1997, ShajiCaves2007} and Fock superpositions~\cite{Turchette2000}, for which dephasing accelerates as $N^2$ and erodes metrological gain.
Intuitively, this can be understood as the rotational phase-space symmetry of Fock states being insufficiently disrupted by small displacements; dephasing of displaced Fock states under the non-Gaussian dephasing channel does not reduce to a simple closed form.
At larger displacements, symmetry-based robustness breaks down and higher-$n$ states incur a penalty, but this regime lies beyond the readout window set by the first zero of $L_n$ for $n \leq 3$ and is not accessed here.
Motional heating couples more strongly to Fock states but is typically orders of magnitude slower than dephasing and is neglected throughout this work~\cite{Brownnutt2015HeatingRateRMP,NoiseHeatingCoherence2016Haffner,GKPSensingValahu2024}.
Above $\tau \sim \qty{1000}{\us}$, dephasing outpaces the gain from longer interrogation and the sensitivity deteriorates, limiting $\tau \lesssim T_2$.
Realizing ultimate sensitivity with quantum enhancement therefore requires operation at short $\tau$ where per-measurement gain persists -- this requirement is met by Qdyne (\Cref{fig:3_qa_qdyne}), which reduces the dependence of gain on $\tau$.

\clearpage
\section{Statistics of the Power and Amplitude Spectra} \label{SI:sec_stats_spectra}
We are interested in the mean and variance of the DFT spectrum calculated from a time-series dataset $\{y_k\}_{k=0}^{N-1}$ consisting of $N$ independent Bernoulli trials with probability $p_k$
\begin{equation}
    \hat{y}_j = \sum_{k=0}^{N-1}{y_k e^{-2i\pi kj/N}}.
\end{equation}
As the DFT is linear, bin components can be decomposed into a deterministic signal component $\mu_{r,i}$ and a random component $n_{r,i}$
\begin{equation}
    \hat{y}_j = \left(\mu_r + i \mu_i\right) + \left(n_r + i n_i \right)
\end{equation}
where subscripts $i,\, r$ denote the real and imaginary terms; the statistics of the background can be obtained by setting $\mu_{i,r} = 0$.

The random components $n_{r,i}$ converge in distribution to normal distributions $\mathcal{N}{(\mu=0,\sigma^2 = N\sigma_y^2/2)}$ via the Lindeberg central limit theorem, where $\sigma_y^2 = \mathrm{Var}{(y_k)}$ is the variance from the time trace, and the factor of $1/2$ follows from equal distribution of noise between real and imaginary components.

\subsection{Expectation and Variance of the Peak Height for a Power Spectrum}
Following~\cite{DegenCASR2017}, the power spectrum is
\begin{equation}
    Y_j = \left|\hat{y}_j\right|^2 = \left(\mu_r + n_r\right)^2 + \left(\mu_i + n_i\right)^2,
\end{equation}
and has expectation
\begin{align} \label{eq:SI_stats_yj_mean}
    \mathbb{E}\left[Y_j\right] &= \left(\mu_r^2 + \mu_i^2\right) + \mathbb{E}\left[n_r^2 + n_i^2\right] + 2\mu_r \mathbb{E}\left[n_r\right] + 2\mu_i \mathbb{E}\left[n_i\right] \nonumber \\
    &= \left| \mu\right|^2 + N\sigma_y^2,
\end{align}
where we have used the fact that $(n_r^2 + n_i^2)/\sigma^2 \sim \chi_2^2$ is chi-squared distributed with two degrees of freedom to obtain $\mathbb{E}\left[n_r^2 + n_i^2\right] = N\sigma_y^2$.\\
Similarly, the variance is
\begin{align} \label{eq:SI_stats_yj_var}
    \mathrm{Var}{\left[Y_j\right]} &= \mathrm{Var}{\left[n_r^2 + n_i^2\right]} + 4\mu_r^2\mathrm{Var}{\left[n_r\right]} + 4\mu_i^2\mathrm{Var}{\left[n_i\right]} \nonumber \\
    &= N^2 \sigma_y^4 + 2\left|\mu\right|^2 N\sigma_y^2.
\end{align}
In addition to the additive noise term contributed by the background, the variance of a peak bin contains an additional term that grows with the peak $\left|\mu\right|^2$ and makes the power spectrum noise heteroscedastic.
In the high-SNR regime $\left|\mu\right|^2 \gg N\sigma_y^2$, the noise of the peak bins can dominate the fit residuals, violating the homoscedasticity assumption underlying ordinary least-squares fitting and yielding incorrect uncertainty estimates for the fitted linecenter (see~\Cref{SI:sec_stats_nllsq}).

\subsection{Expectation and Variance of the Peak Height for an Amplitude Spectrum}
The statistics of the amplitude spectrum $\left|\hat{y}_j\right| = \left| (\mu_r + i\mu_i) + (n_r + i n_i) \right|$ are easily found by noticing that $\left|\hat{y}_j\right|$ is the modulus of a complex normal random variable and is thus Rice-distributed, i.e. $\left|\hat{y}_j\right| \sim \mathrm{Rice}{\left(\nu = \left|\mu\right|, \sigma^2=N\sigma_y^2/2\right)}$, where $\nu$ is the Rician noncentrality parameter.
The expectation and variance are
\begin{equation}
    \mathbb{E}{\left[\left|\hat{y}_j\right|\right]} = \sqrt{\frac{\pi N \sigma_y^2}{4}} L_{\frac{1}{2}}{\left(-\frac{\left|\mu\right|^2}{N\sigma_y^2}\right)}, \qquad
    \mathrm{Var}{\left[\left|\hat{y}_j\right|\right]} = N\sigma_y^2 + \left|\mu\right|^2 - \frac{\pi N \sigma_y^2}{4} L_{\frac{1}{2}}{\left(-\frac{\left|\mu\right|^2}{N\sigma_y^2} \right)}^2,
\end{equation}
where $L_{\frac{1}{2}}{(x)}$ is the Laguerre polynomial of order $1/2$.
For $x \gg 1$~\cite{Gudbjartsson1995}, corresponding to the high-SNR limit,
\begin{equation}
    L_{\frac{1}{2}}\!{(x)} = \sqrt{\frac{4x}{\pi}} \left(1 + \frac{1}{4x}\right),
\end{equation}
recovering
\begin{equation}    \label{eq:SI_stats_yj_mean_variance_sqrt}
    \mathbb{E}{\left[\left|\hat{y}_j\right|\right]} \approx \left|\mu\right| \left(1 + \frac{N\sigma_y^2}{4\left|\mu\right|^2}\right), \qquad
    \mathrm{Var}{\left[\left|\hat{y}_j\right|\right]} \approx \frac{N\sigma_y^2}{2}.
\end{equation}
In contrast to~\cref{eq:SI_stats_yj_var}, the variance of the peak is independent of its height.
For $\mu = 0$, the Rice distribution reduces to a Rayleigh distribution with $\sigma^2 = N\sigma_y^2/2$, such that the background is
\begin{equation}    \label{eq:SI_stats_rayleigh_var}
    \mathbb{E}{\left[\textrm{background}\right]} = \sqrt{\frac{\pi N \sigma_y^2}{4}}, \qquad
    \mathrm{Var}{\left[\textrm{background}\right]} = \frac{4-\pi}{2} \, \frac{N\sigma_y^2}{2}.
\end{equation}
Though the amplitude spectrum is not strictly homoscedastic, the ratio of background to peak variance is fixed and small at $2-\pi/2 \approx 0.43$, enabling robust least-squares fitting.

\clearpage
\section{Uncertainty of the Fitted Linecenter} \label{SI:sec_stats_nllsq}
Contrary to the convention of fitting a Lorentzian $L = \frac{h}{1+u^2}$ to the power spectrum, we fit $\sqrt{L}$ to the amplitude spectrum (see~\Cref{SI:sec_stats_spectra}) throughout this work to ensure agreement of the reported uncertainty of the fitted linecenter $\sigma_f$ with its true uncertainty.\\
When fitting a general model $\hat{y}{(x)}$ with parameters $\{\beta_i\}$ to a set of measurements $\{y_i\}$, the variance of $\beta_i$ is estimated as
\begin{equation}
    \sigma_i^2 = \sigma_{\textrm{res}}^2 \left[\left(J^T J\right)^{-1}\right]_{ii},
\end{equation}
where $J$ is the Jacobian matrix, and $\sigma_{\mathrm{res}}^2$ is the variance of the residuals
\begin{equation}
    J_{ij} = \left.\frac{\partial y}{\partial \beta_j} \right|_{x_i}, \qquad
    \sigma_{\mathrm{res}}^2 = \frac{1}{N-d} \sum_{i}^{N}{\left(y_i - \hat{y}{(x_i)}\right)^2},
\end{equation}
and $d$ is the number of fit parameters.
The matrix term $\left(J^T J\right)_{ij}$ forms a Riemann sum, and for $N \gg1$ can be approximated by the corresponding integral:
\begin{align}    \label{eq:SI_stats_normalMatrix_general}
    \left(J^T J\right)_{ij} &= \sum_{k}^{N}{ \left.\frac{\partial y}{\partial \beta_i}\right|_{x_k} \left.\frac{\partial y}{\partial \beta_j}\right|_{x_k}}
        = \frac{1}{\Delta x}  \sum_{k}^{N}{ \left.\frac{\partial y}{\partial \beta_i}\right|_{x_k} \left.\frac{\partial y}{\partial \beta_j}\right|_{x_k}} \, \Delta x \nonumber \\
    &\approx  \frac{1}{\Delta x} \int_{x_1}^{x_N}{\frac{\partial  y(x)}{\partial \beta_i} \frac{\partial y(x)}{\partial \beta_j} \, dx}.
\end{align}

\subsection{Least-squares uncertainty for Lorentzian models}
We consider Lorentzian models of the form
\begin{equation}    \label{eq:SI_stats_lorentzian_general}
    L{(x)} = A \, \frac{1}{\Gamma^2 + \left(x-f\right)^2} + C,
\end{equation}
where $A$ is the amplitude, $\Gamma$ is the linewidth, $f$ is the linecenter, and $C$ is an offset; this parameterization ensures $A$ is independent of $\Gamma$ and thus of measurement time.
This can be nondimensionalized as
\begin{equation}    \label{eq:SI_stats_lorentzian_nondimensionalized}
    L{(x)} = \frac{h}{1 + u^2} + C,
\end{equation}
where $h = A/\Gamma^2$ and $u = (x-f)/\Gamma$.
The relevant partial derivatives are
\begin{align}
    \partial_{A}{L} = \frac{1}{\Gamma^2} \frac{1}{1+u^2},
    && \partial_\Gamma{L} = -\frac{2h}{\Gamma} \frac{1}{\left(1+u^2\right)^2},
    && \partial_f{L} = \frac{2h}{\Gamma} \frac{u}{\left(1+u^2\right)^2},
    && \partial_C{L} = 1.
\end{align}
If the Lorentzian peak is far from the integration bounds $\lvert x_{1,N} - f \rvert \gg \Gamma$, the integrands in~\cref{eq:SI_stats_normalMatrix_general} are peak-dominated and vanish as $x \rightarrow \pm \infty$, and the limits can be approximated as $\pm\infty$.
Substituting $u = (x-f)/\Gamma$ to center the peak at the origin and using the Fourier bin spacing $\Delta x = 1/T$,~\cref{eq:SI_stats_normalMatrix_general} becomes
\begin{equation}
    \left(J^T J\right)_{ij} \approx T \Gamma \int_{-\infty}^{\infty}{\frac{\partial  y(u)}{\partial \beta_i} \frac{\partial y(u)}{\partial \beta_j} \, du}.
\end{equation}

\subsection{Lorentzian fits to the power spectrum}
We use a fit model $y=L(x)$, i.e. the straightforward Lorentzian.
Evaluation of~\cref{eq:SI_stats_normalMatrix_general} gives
\begin{equation}
    J^T \!J = \begin{pmatrix}
        \frac{\pi T}{2\Gamma^3} & -\frac{3A\pi T}{4\Gamma^4} & 0 & \frac{\pi T}{\Gamma} \\
        -\frac{3A\pi T}{4\Gamma^4} & \frac{5A^2 \pi T}{4\Gamma^5} & 0 & -\frac{A\pi T}{\Gamma^2} \\
        0 & 0 & \frac{A^2\pi T}{4\Gamma^5} & 0\\
        \frac{\pi T}{\Gamma} & -\frac{A\pi T}{\Gamma^2} & 0 & N
    \end{pmatrix},
\end{equation}
such that the diagonal elements of its inverse are
\begin{equation}
    \left[\left(J^T \!J\right)^{-1}\right]_{ii} = \left(
        \frac{4\Gamma^3}{\pi T} \frac{5N-4\pi T\Gamma}{N - 4\pi T \Gamma}, \qquad
        \frac{8\Gamma^5}{A^2 \pi T} \frac{N - 2\pi T \Gamma}{N-4\pi T \Gamma},  \qquad
        \frac{4 \Gamma^5}{A^2 \pi T},  \qquad
        \frac{1}{N - 4\pi T \Gamma}
    \right).
\end{equation}
The variance of the estimated linecenter $\sigma_f^2$ is then
\begin{equation}
    \sigma_f^2 = \sigma_{\mathrm{res}}^2 \frac{4 \Gamma^5}{A^2 \pi T}.
\end{equation}
We are interested in the regime where the intrinsic linewidth is poorly resolved by the DFT frequency bins, i.e. the peak linewidth is simply $\Gamma = 1/T$.
Here, $\sigma_{\mathrm{res}}^2$ is comprised of the variances from the background and the peak -- from~\cref{eq:SI_stats_yj_var}, these are~\cite{DegenCASR2017}
\begin{equation}
    \mathrm{Var}{(\mathrm{background})} = N^2 \sigma_y^4, \qquad
    \textrm{Var}{(\textrm{peak})} = N^2 \sigma_y^4 + 2N\sigma_y^2 \left|\mu\right|^2,
\end{equation}
where $\sigma_y^2$ is the variance of a point in the time trace, and $\left|\mu\right|^2 = A/\Gamma^2$ is the peak height.
The residual term is then
\begin{align}
    \sigma_{\mathrm{res}}^2 &\approx \frac{1}{N_\mathrm{fit}} \left(N_\mathrm{fit}\times \mathrm{Var}{(\textrm{background})} + \mathrm{Var}{(\textrm{peak})} \right) \nonumber \\
    &= N \sigma_y^2 \left[ N \sigma_y^2\left( 1 + \frac{1}{N_\mathrm{fit}} \right)
        + \frac{1}{N_\mathrm{fit}} \frac{2A}{\Gamma^2} \right].
\end{align}
Finally, $\sigma_f^2$ is
\begin{align}
    \sigma_f^2 &\approx \frac{4 \Gamma^5}{A^2 \pi T} \: N \sigma_y^2 \left(
        N \sigma_y^2\left( 1 + \frac{1}{N_\mathrm{fit}} \right)
        + \frac{1}{N_\mathrm{fit}} \frac{2A}{\Gamma^2} \right) \nonumber \\
    &\approx \frac{4 \sigma_y^2 }{A^2 \pi T_s} \left(
        \frac{1}{T^4} \frac{\sigma_y^2}{T_s} + \frac{1}{T^3} \frac{2A}{N_\mathrm{fit}} \right)
\end{align}
where we have used $\Gamma = 1/T$, $N = T/T_s$, and $T_s$ is the sample period.

For large fit windows $N_\mathrm{fit} \gg 2A T^2/(N\sigma_y^2)$, or if all spectrum points $N_\mathrm{fit} = N$ are fitted, the background term dominates and $\sigma_f \propto T^{-2}$, exceeding the Cram\'er-Rao scaling $\sigma_f \propto T^{-3/2}$~\cite{RifeBoorstyn1974}; for a small window, the peak term dominates and the correct scaling is recovered.
This is due to the heteroscedasticity of~\cref{eq:SI_stats_yj_var}, in which $\sigma_\mathrm{res}^2$ reflects only the background and underestimates the true uncertainty by an SNR factor $\left|\mu\right|/\sqrt{N\sigma_y^2/2}$.
This creates a metrological ambiguity as the reported precision can be tuned almost arbitrarily by choice of $N_\mathrm{fit}$, and relative gains between datasets are overestimated by a factor of the SNR.

\subsection{Square-root Lorentzian fits to the amplitude spectrum}
We consider the model
\begin{equation}
    y{(x)} = \sqrt{A \, \frac{1}{\Gamma^2 + \left(x-f\right)^2} + C},
\end{equation}
where $A$ is an amplitude, $\Gamma$ is the linewidth, $f$ is the linecenter, and $C$ is an offset.
The uncertainty in the linecenter is then
\begin{equation}
    \sigma_f^2 = \sigma_{\textrm{res}}^2 \frac{8\Gamma^3}{A\pi T}.
\end{equation}
Using~\cref{eq:SI_stats_yj_mean_variance_sqrt,eq:SI_stats_rayleigh_var}, the residual term is thus
\begin{align}
    \sigma_{\mathrm{res}}^2 &\approx \frac{1}{N_\mathrm{fit}} \left(N_\mathrm{fit}\times \mathrm{Var}{(\textrm{background})} + \mathrm{Var}{(\textrm{peak})} \right) \nonumber \\
    &= \frac{N \sigma_y^2}{2} \left( \frac{4-\pi}{2} + \frac{1}{N_\mathrm{fit}}\right),
\end{align}
and consequently,
\begin{align}   \label{eq:SI_stats_sqrtL_final}
    \sigma_f^2 &= \frac{8\Gamma^3}{A\pi T} \, \frac{N \sigma_y^2}{2} \left( \frac{4-\pi}{2} + \frac{1}{N_\mathrm{fit}}\right) \nonumber \\
    &= \frac{4\sigma_y^2}{A\pi T_s} \, \frac{1}{T^3} \left( \frac{4-\pi}{2} + \frac{1}{N_\mathrm{fit}}\right),
\end{align}
where we have again used $\Gamma = 1/T$, $N = T/T_s$, and $T_s$ is the sample period.
\Cref{eq:SI_stats_sqrtL_final} thus recovers the Cram\'{e}r--Rao scaling $\sigma_f \propto T^{-3/2}$~\cite{RifeBoorstyn1974} and is independent of $N_\mathrm{fit}$ for $N_\mathrm{fit} \gg 1$.

\clearpage
\section{Estimating metrological gain} \label{SI:sec_gain_estimation}
\subsection{Estimating gain from the Fisher Information}
The Standard Quantum Limit in~\cref{eq:SI_sql_qdyne_1} assumes that every measurement saturates the quantum Fisher information $\mathfrak{F} = 4(2n+1)$.
A more realistic estimate uses the Fisher information $F(\alpha)$ obtained from fits to the data in~\Cref{fig:3_qa_qdyne}(a,b).
From~\cref{eq:SI_sql_qdyne_1}, the Fisher information is
\begin{equation}   \label{eq:SI_fi_gain_estimate}
     F_\textrm{tot} = \sum_{j=0}^{N}{F(\alpha_j) \left(\frac{\partial \alpha_j}{\partial \delta}\right)^2}
     = \sum_{j=0}^{N}{F(\alpha_0 \sin{(\delta t_j)}) \left( t_j \alpha_0 \cos{(\delta t_j)} \right)^2},
\end{equation}
where the displacement is sinusoidal in time, $\alpha(t) = \alpha_0\sin{(\delta t)}$.
For $\delta t_N \gg 1$, this sum is proportional to the phase average
\begin{equation}    \label{eq:SI_fi_gain_avg_final}
    \langle F\rangle_\phi = \frac{1}{2\pi} \int_{0}^{2\pi}{\alpha_0^2 \: F\!\left(\alpha_0 \sin{\phi}\right) \cos^2\!{\phi}  \; d\phi}.
\end{equation}
This result is used to predict the metrological performance of different Fock states using independently characterized experimental parameters (e.g.~\Cref{fig:3_qa_qdyne}(b), following~\cref{eq:fock_prob_fit}).

\subsection{Extracting $\alpha_0$ from the Qdyne measurement record}    \label{SI:sec_gain_estimation_alpha0_only}
For an oscillating displacement $\alpha = \alpha_0\cos\phi$, the record is described by~\cref{eq:fock_prob_fit} with $S_n(\alpha) = 1 - e^{-\alpha^2} L_n\!{(\alpha^2)}^2$, where $n$ is the Fock state.
Its Fourier coefficients are
\begin{equation}
    c_0{(\alpha_0)} = \frac{1}{2\pi} \int_{0}^{2\pi}{S_n{(\alpha_0 \cos{(\phi)})} \, d\phi}, \qquad
    c_{2m} = \frac{1}{\pi} \int_{0}^{2\pi}{S_n{(\alpha_0 \cos{(\phi)})} \cos\!{(2m\phi)} \, d\phi},
\end{equation}
where odd harmonics do not contribute as $S_n$ is even in $\alpha$.
The spectral peak heights are then
\begin{equation}
    \mathrm{peak}_{2m} = \frac{N A \,c_{2m}{(\alpha_0)}}{2},
\end{equation}
where $\mathrm{peak}_{2m}$ is the peak height, $N$ is the number of dataset points, and $A$ is a fit parameter from~\cref{eq:fock_prob_fit} representing signal contrast.
Knowledge of $A$ (from e.g. calibration, as in~\Cref{fig:3_qa_qdyne}(b)) can thus be used to extract $\alpha_0$ from the Qdyne spectrum, which is invertible for $\alpha_0$ within the monotonic range of $c_{2m}$ used here.
As sensitivity to $\alpha_0$ is lower for $2m \geq 4$ harmonics, and the $c_0$ term is sensitive to drifts in readout background (e.g. changes in SPAM fidelity), we extract $\alpha_0$ via the $c_{2}$ peak.

\clearpage
\section{Quantum-Enhanced Gain and Realized Displacements} \label{SI:sec_gain_disparity}
\subsection{Calibrating $\alpha_0$}
Following acquisition of each dataset in~\Cref{fig:3_qa_qdyne}(c), a short, separate Qdyne dataset of $10^{5}$ samples was acquired to calibrate the displacement $\alpha_0$ experimentally.
A DFT is then applied to the calibration dataset to extract $\alpha_0$ following~\Cref{SI:sec_gain_estimation_alpha0_only}.
This determines $\alpha_0$ under conditions as close as possible to those during the measurement, as experimental factors such as duty cycle can shift the realized $\alpha_0$.
Standard calibrations (e.g. linescans) carry high uncertainty at the small displacements $\alpha_0 \sim 0.3$--$0.4$ here and are instead performed at $\alpha \sim 1$ then extrapolated, resulting in systematic shifts from amplifier and DDS nonlinearity and pulse distortion.
This in-situ calibration is also robust to secular frequency drifts, as our implementation of Qdyne does not depend explicitly on the secular frequency.

The calibrated displacements are $\alpha_0 = 0.32(2) ,\, 0.31(2) ,\, 0.40(2) ,\, 0.38(2)$ for the $n=0,1,2,3$ datasets.
The marked difference between the $n=0,1$ and $n=2,3$ datasets, which were acquired on different days, reflects system drift between acquisitions, such as shimming of the trap potential, which alters the local field gradient, and trap impedance changes due to hardware changes and lab weather-induced drift of component values.
A uniform uncertainty $\delta\alpha_0 = 0.02$ is assigned to all displacements, obtained from a separate measurement of the long-term drift rate combined with the statistical uncertainty from the spectral noise floor.

\subsection{Measuring raw gains}
Metrological gains in~\Cref{fig:3_qa_qdyne}(c) are extracted from linear fits $\log \sigma_\delta = \log b - 1.5 \log T$ to the log-transformed data (\Cref{tab:SI_3c_scaling_values}(a)).
The measured gain of Fock state $n$ relative to $n=0$ is
\begin{equation}    \label{eq:SI_gain_measured}
    g_n^{\mathrm{meas}} = 20\log_{10}\!{\left( \frac{b_{n=0}}{b_{n}} \right)}
\end{equation}
with uncertainty propagated from $1\sigma$ uncertainties in $b_0,b_n$; resulting values are tabulated in~\Cref{tab:SI_gain_realized}.

\subsection{Predicting gains}
Gains are predicted by comparing the precision obtained to that with the $n=0$ state, expressed as
\begin{equation}    \label{eq:SI_gain_pred_realized}
    g_n^{\mathrm{pred}}\!{\left(\alpha_0^{(n)}, \alpha_0^{(0)}\right)}
    = 10 \log_{10}\!{\left( \frac{\langle F \rangle_\phi^{(n)}{\left(\alpha_0^{(n)}\right)}}{\langle F \rangle_\phi^{(0)}{\left(\alpha_0^{(0)}\right)}} \right)},
\end{equation}
following~\cref{eq:SI_fi_gain_avg_final}.
This requires values for $A_n,\,C_n,\,\alpha_0$ from~\Cref{tab:SI_ADJCDJ_fig3a,tab:SI_3c_scaling_values}; results are tabulated in~\Cref{tab:SI_gain_realized} with uncertainties obtained by first-order propagation.
These gains do not solely represent the quantum-enhanced contribution from the Fock state, and contain contributions from the displacement through $\sigma_\delta^n \propto 1/\alpha_0$ and $F(\alpha_0 \sin\phi)$.

\subsection{Displacement-corrected gain}
To isolate the contribution of the Fock state, we define the displacement contribution
\begin{equation}    \label{eq:SI_gain_delta}
    \Delta_n = g_n^{\mathrm{pred}}\!{\left(\alpha_0^{(n)}, \alpha_0^{(0)}\right)} - g_n^{\mathrm{pred}}\!{\left(\alpha_0^{(0)}, \alpha_0^{(0)}\right)},
\end{equation}
i.e. the excess gain predicted for the $n$th Fock state due to the difference in displacement, and consequently, the displacement-corrected gain
\begin{equation}    \label{eq:SI_gain_corr}
    g_n^{\mathrm{corr}} = g_n^{\mathrm{meas}} - \Delta_n,
\end{equation}
i.e. the gain achieved by each Fock state.
Displacement-corrected gains (\Cref{tab:SI_gain_realized}) agree with predictions at the common $n=0$ displacement within uncertainties.

\begin{table*}[h!]
    \centering
    \begin{tabular}{|c|c||c|c|c||c|}
        \hline
        & \textbf{Measured} & \multicolumn{3}{|c||}{\textbf{Calculated}} & \textbf{Displacement-Corrected} \\
        \hline
         & $g_n^{\mathrm{meas}}$ (dB) & $g_n^{\mathrm{pred}}(\alpha_0^{(n)}, \alpha_0^{(0)})$ (dB) & $g_n^{\mathrm{pred}}(\alpha_0^{(0)}, \alpha_0^{(0)})$ (dB) & $\Delta_n$ (dB) & $g_n^{\mathrm{corr}}$ (dB) \\
        \hline
        $\ket{n=1}$ & 4.55(12) & 4.6(1.4) & 4.63(7) & -0.1(1.2) & 4.7(1.2) \\
        $\ket{n=2}$ & 8.90(12) & 9.1(1.2) & 6.35(7) & 2.7(1.0) & 6.2(1.0) \\
        $\ket{n=3}$ & 9.17(11) & 8.5(1.2) & 6.32(7) & 2.1(1.0) & 7.1(1.0) \\
        \hline
    \end{tabular}
    \caption{Measured, predicted, and displacement-corrected gains over the $n=0$ state for Qdyne measurements in~\Cref{fig:3_qa_qdyne}(c).
    Measured gains are calculated directly from the dataset following~\cref{eq:SI_gain_measured}.
    Predictions use~\cref{eq:SI_fi_gain_avg_final} with parameters from~\Cref{fig:3_qa_qdyne}(b) and calibrated displacements $\alpha_0^{(n)} = 0.32,\,0.31,\,0.40,\,0.38$ for $n=0,1,2,3$.
    Displacement-corrected gains ($g_n^{\mathrm{corr}} ,\, \Delta_n$) are evaluated using~\cref{eq:SI_gain_delta,eq:SI_gain_corr} at common $\alpha_0^{(0)} = 0.32$.
    Uncertainties are $1\sigma$.
    \label{tab:SI_gain_realized}
    }
\end{table*}